\documentclass[11pt,twoside,letterpaper]{article}
\usepackage{times}
\usepackage{fancyhdr}
\usepackage{amssymb}
\usepackage{amsmath}
\usepackage{graphicx}
\usepackage{color}
\usepackage[cmtip,arrow]{xy}
\usepackage{pb-diagram,pb-xy}
\usepackage{epstopdf}	


\makeatletter 
\def\cleardoublepage{\clearpage\if@twoside \ifodd\c@page\else%
    \hbox{}%
    \thispagestyle{empty}%
    \newpage%
    \if@twocolumn\hbox{}\newpage\fi\fi\fi} 
\makeatother

\def\figurename{Figure}
\makeatletter
\renewcommand{\fnum@figure}[1]{\figurename~\thefigure.}
\makeatother

\def\tablename{Table}
\makeatletter
\renewcommand{\fnum@table}[1]{\bfseries\tablename~\thetable.}
\makeatother

\def\A{\mathbb{A}} 
\def\C{\mathbb{C}} 
\def\F{\mathbb{F}} %
\def\Z{\mathbb{Z}}
\def\S{\mathbb{S}}
\def\T{\mathbb{T}}
\def\R{\mathbb{R}}

\def\e{\mathrm{e}}
\def\esp{\Pi}
\def\U{\mathrm{U}}
\def\sG{\mathrm{G}} 
\def\sg{\mathrm{g}} 
\def\sGN{{N_\sG}} 
\def\sGloc{\sG_{\mathrm{loc}}} 
\def\Time{\mathcal{T}} 
\def\iG{\Gamma} 
\def\ig{\mathbf{\gamma}} 
\def\iGN{{N_\iG}} 
\def\W{\mathrm{W}} 
\def\X{\mathrm{X}} 
\def\x{\mathrm{x}} 
\def\XN{{N_\X}} 
\def\iGX{\iG^{\X}}
\def\PF{P}
\def\Partransport{\pi}
\def\Rtimes{\rtimes}
\def\Times{\times}
\def\SpaceTime{\mathcal{M}} 
\def\wS{\mathcal{S}} 
\def\ws{\mathbf{s}} 
\def\wSN{{N_\wS}} 
\def\Evoset{\mathcal{E}} 
\def\wG{\mathcal{W}} 
\def\wg{\mathrm{w}} 
\def\wGN{{N_\wG}} 
\def\id{\mathbf{1}} 
\def\ls{\mathbf{\sigma}} 
\def\lS{\Sigma} 
\def\lSN{{N_\lS}} 
\def\lSX{\lS^{\X}}
\def\lSXN{\left|\lSX\right|}

\def\Tr{\mathrm{Tr}}

\def\Hspace{\mathcal{H}}
\def\lstep{2} 
\def\rstep{1} 
\def\lprob{p_\lstep} 
\def\rprob{p_\rstep} 
\def\veloc{v} 
\def\transmatr{\mathrm{S}}
\def\irrep{\Delta}
\def\PP{\widetilde{P}}
\def\regrep{\mathrm{P}}
\def\Repbare{\rho}
\def\Charbare{\chi}
\def\CL{\emph{\textbf{Conway's Life}}}
\def\RCL{R^\delta_{CL}}
\def\PCL{P_{CL}}
\newcommand{\Alt}[1]{\mathrm{Alt}\!\left(#1\right)} 
\newcommand{\Cgr}[1]{\Z_{#1}} %
\newcommand{\Dih}[1]{\mathrm{D}_{#1}} %
\newcommand{\Perm}[1]{\mathrm{Sym}\left(#1\right)} 
\newcommand{\Rep}[1]{\Repbare\left(#1\right)} %
\newcommand{\RepH}[1]{\Repbare^\dagger\left(#1\right)} %
\newcommand{\Char}[1]{\Charbare\left(#1\right)} 
\newcommand{\rbra}[1]{\left(#1\right)} 
\newcommand{\set}[1]{\left\{#1\right\}} 
\newcommand{\vect}[1]{\left(#1\right)} 
\newcommand{\inrb}[1]{\left(#1\right)} 
\def\wbrl{\left(}
\def\wbrr{\right)}
\def\wmult{} %
\newcommand{\welem}[2]{\wbrl{#1},~{#2}\wbrr} %
\newcommand{\esymm}[2]{\esp_{#1}\inrb{#2}}

\begin{document}
\title{
{\begin{flushleft} 
\vskip 0.45in
{\normalsize\bfseries\textit{Chapter~7}} 
\end{flushleft} 
\vskip 0.45in
\bfseries\scshape Structural and Symmetry Analysis of Discrete Dynamical Systems}}
\author{
\bfseries\itshape Vladimir V. Kornyak\thanks{E-mail address: kornyak@jinr.ru}\\
Joint Institute for Nuclear Research\\
141980 Dubna, Russia}

\date{}
\maketitle
\thispagestyle{empty}
\setcounter{page}{1}
\thispagestyle{fancy}
\fancyhead{}
\fancyhead[L]{In: Cellular Automata \\ 
Editor: Thomas M. Li} 
\fancyhead[R]{ISBN: 978-1-61761-592-4\\
\copyright~2010 Nova Science Publishers, Inc.}
\fancyfoot{}
\renewcommand{\headrulewidth}{0pt}

\begin{abstract}
\noindent
To study 
discrete dynamical systems of different 
types --- {deterministic}, {statistical} and {quantum} 
--- we develop various approaches.%
We introduce the concept of a system of discrete relations 
on an abstract simplicial complex and develop  algorithms 
for~ analysis of compatibility and construction of canonical 
decompositions of such systems. To illustrate these techniques 
we describe their application to some cellular automata.
Much attention is paid to study symmetries of the systems.
In the case of deterministic  systems, we reveal some important
relations between symmetries and dynamics. 
We demonstrate that moving soliton-like structures 
arise inevitably in deterministic dynamical system whose symmetry 
group splits the set of states into a finite number of group orbits.
We develop algorithms and programs exploiting discrete symmetries to study
{microcanonical ensembles} and search {phase transitions} in mesoscopic 
lattice models.
We propose an approach to quantization of discrete systems
based on introduction of gauge connection with values in unitary 
representations of {finite}  groups 
--- the elements of the connection are interpreted as amplitudes of quantum transitions.
We discuss properties of a quantum description of finite systems. In particular, we demonstrate
that a finite quantum system can be embedded into a larger classical system.  
Computer algebra and computational group theory methods were useful tools in our study. 
\end{abstract}
\noindent 
\textbf{Keywords:}
discrete relations, cellular automata, symmetries of discrete systems,
discrete gauge principle, quantization, computer algebra
\par
\vspace{.08in}
\noindent
{\textbf MSC 2000:} 37A, 37B, 68W, 81R, 81T.
\section{Introduction}
There are many reasons --- physical, mathematical, and conceptual 
---  to study discrete structures. Discrete systems are important in
applications --- \textit{nanostructures}, for example, by
their nature are discrete, not continuous, formations. 
From a fundamental point of view, there are many philosophical
and physical arguments that discreteness better describes 
physics%
\footnote{Of course, the question of ``whether the real world is discrete or continuous''
and even ``finite or infinite'' is rather
\emph{metaphysical}, i.e., neither empirical observations nor logical arguments can validate
one of the two adoptions --- this is a matter of belief or taste.}
at small distances than continuity which arises only as 
approximation or as a logical limit in
considering large collections of discrete structures.
As a recent development, let us mention much-discussed E. Verlinde's thermodynamic
(entropic) derivation \cite{Verlinde} of gravity and Newton's law of inertia from G. 
't Hooft's \textit{holographic principle} \cite{tHooft}. The holographic principle
conjectures that it is possible to describe physical events in a three-dimensional
volume fully by a theory on its temporally varying two-dimensional boundary
--- \emph{holographic screen} --- containing \emph{finite} number of
discrete degrees of freedom. Entropy of these degrees of freedom, i.e., 
number of bits $N$, is proportional to the area $A$ of the screen:
$N=\frac{\textstyle{Ac^3}}{\textstyle{G\hbar}}$.%
\footnote{In theories with \emph{emergent space}
this relation may be used as definition of area: each fundamental 
bit occupies by definition one unit of area.} In more speculative sense, the whole universe 
is a finite two-dimensional  information structure on the \emph{cosmological horizon}, 
and observable three dimensions are only an effective description at macroscopic scales 
and at low energies.
Verlinde shows that the laws of Newton and the Einstein equations 
come out directly and unavoidably from the holographic principle.
The gravity appears to be an \emph{entropic force} arising in systems with many 
degrees of freedom by the statistical tendency to increase its entropy --- 
like osmosis or elasticity of polymers.
Verlinde derived his results combining holography
 $\left({N=\frac{\textstyle{Ac^3}}{\textstyle{G\hbar}}}\right)$, 
 the equipartition rule  (assumption on even distribution of energy over $N$ bits),
1st law of thermodynamics $\left({dE=TdS-Fdx}\right)$ and several 
additional standard relations.
To introduce thermodynamics, i.e., to construct 
\emph{canonical partition function}, 
there is no need to know details of microscopic dynamics. 
It suffices to know about \emph{energy} and number of states.
Of course, the fundamental problem
about laws governing bit dynamics on holographic screens remains unsolved.
Since Planck scales are experimentally unavailable --- the Planck length is about 
$10^{-35}$ meters, i.e., far below the spacial resolution 
of particle accelerators (nowadays about $10^{-18}$ meters) 
--- the construction and study 
of various discrete dynamical models is one of the possible approaches.
 
\pagestyle{fancy}
\fancyhead{}
\fancyhead[EC]{Vladimir V. Kornyak}
\fancyhead[EL,OR]{\thepage}
\fancyhead[OC]{
Analysis of Discrete Dynamical Systems}
\fancyfoot{}
\renewcommand\headrulewidth{0.5pt}

In this chapter we consider three types of  discrete dynamical
systems: deterministic, mesoscopic statistical and quantum.
 
We begin with a general discussion of dyscrete dynamical systems. The most fundamental
concepts are a discrete time and a set of states evolving in the time. 
A space is considered as a derived concept providing the set of states with the specific structure
of a set of functions on the points of space with values in some set of local states. 
We give an illustration of how a space-time may arise in simple models of discrete dynamics.
Then we discuss symmetries of space and local states and how these symmetries can 
be combined into a single group of symmetries of the system as a whole.
\par
We introduce the concept of a system of 
\textit{discrete relations on an abstract simplicial complex} \cite{KornyakOC,KornyakDR}, 
and explain how any system of
discrete relations
--- subsets of Cartesian products of finite sets --- acquires the structure of an abstract
simplicial complex. This general concept
covers many discrete mathematical structures. In particular,  it
can be considered as generalization of cellular automata or as a set-theoretical analog of
systems of polynomial equations --- if all factors of the Cartesian product are sets 
with the same number of elements and this number is  \textit{prime power}, than any relation can be
expressed by polynomial equation. 
We describe algorithms 
for analysing \textit{compatibility} and constructing \textit{canonical decompositions}
of discrete relations. As an illustration, we give results of application of the algorithms
to some cellular automata, namely, Conway's automaton \textit{Game of Life} and Wolfram's
\textit{elementary cellular automata}. For many of the latter automata 
the canonical decomposition allows to obtain either general solutions in closed form or
important information on their global behavior.
\par
Symmetry is a property of fundamental importance for any mathematical or physical structure.
Many real world discrete systems, e.g., carbon nanostructures like graphenes and fullerenes, 
are highly symmetric formations.
Symmetries play essential role in the dynamics of the systems.
In this chapter we consider connection
between symmetries of discrete dynamical systems on graphs ---
1-dimensional simplicial complexes --- and their dynamics  \cite{KornyakCASC06,Kornyak08}.
In the case of \emph{deterministic dynamical systems}, such as cellular
automata, non-trivial connections between the lattice symmetries and
dynamics are revealed. In particular, we show that formation of moving
soliton-like structures --- typical examples are ``spaceships'' in cellular 
automata --- is a direct result of the existence of non-trivial symmetry.
\par%
We developed also algorithms exploiting symmetries for computing microcanonical 
partition functions and for searching phase transitions in \emph{mesoscopic lattice models}.
\par
We consider a class of discrete dynamical models allowing quantum
description \cite{KornyakCASC09}. Our approach to quantization consists 
in introduction of
gauge connection with values in unitary representation
(not necessarily 1-dimensional) of some group of \emph{internal symmetries} 
 --- the elements
of the connection are interpreted as amplitudes of quantum transitions.
The standard quantization is a special case of this construction ---
Feynman's path amplitude $\e^{i\int{Ldt}}$ can be interpreted as
parallel transport with values in (1-dimensional) fundamental
representation $\U(1)$ of the group of phase transformations. 
For discrete systems it is natural 
to take a \textit{finite} group as the \textit{quantizing} group, in this case all  
manipulations --- in contrast to the standard quantization --- remain within the framework
of constructive discrete mathematics requiring no more than the ring of
\emph{algebraic integers}  (and sometimes the quotient field of this ring). 
On the other hand, the standard quantization can be approximated by taking 
1-dimensional representations of large enough finite groups. 
\par 
Any approach to quantization leads ultimately to unitary operators acting on 
a Hilbert space. We discuss peculiarities of quantum description of finite 
systems, under the assumption that the operators describing quantum behavior  
are elements of unitary representations of finite groups. We show that in this 
case any quantum problem can be embedded into a classical one with a larger
space of representation.
\par
Computer algebra and computational group theory \cite{Holt} methods 
turned out to be quite useful tools in our study of discrete systems.
\section{Discrete Dynamics}
Generally, \emph{discrete dynamical system} is a set $\wS=\set{\ws_1,\ldots,\ws_{\wSN}}$ of 
distinguishable states evolving in \emph{discrete time} 
$t\in\Time\cong\Z=\set{\ldots,-1,0,1,\ldots}$, i.e., \emph{evolution} or \emph{history}
is an element of the set $\Evoset=\wS^\Time$.
\emph{Dynamics} is determined by some \emph{evolution rule} connecting 
the current state
$s_t\in\wS$ of the system with its prehistory $s_{t-1},$ $s_{t-2},$ $s_{t-3},\ldots$
Different types of evolution rules are possible. We shall consider here the following types of
discrete dynamics. 
\begin{itemize}
	\item 
Evolution rule of \emph{deterministic dynamical system} is 
a \emph{functional relation}. 
This means that the current state is a function of the prehistory:
\begin{equation}
	s_t=F\left(s_{t-1}, s_{t-2}, s_{t-3}, \ldots\right).
	\label{funcrel}
\end{equation}
{Cellular automaton} is a typical example of deterministic dynamical system.
	\item 
\emph{Statistical lattice model} is a sort of non-deterministic dynamical system.
This is a special case of \emph{Markov chain}. In statistical lattice model transition 
from one state to any other is possible with probability controlled by a Hamiltonian.
	\item 
\emph{Quantum system} is another important type of non-deterministic dynamical system.
The probabilities of transitions between states are expressed 
in terms of complex-valued transition amplitudes.
\end{itemize}
\par
Symmetries play an important --- central in the case of quantum systems --- role in 
dynamical systems.
So we assume the existence of a non-trivial group $\wG=\left\{\wg_1=\id, \wg_2,\right.$
$\ldots,$
$\left.\wg_{\wGN}\right\}$\footnote{We denote the identity elements by $\id$ for~ all groups 
throughout this chapter.}
acting on the set of states $\wS$:  $\wG\leq\Perm{\wS}$.
Action of the group $\W$ splits the set of states $\wS$ 
	into \emph{orbits} of different sizes: 
	$\wS=\bigsqcup\limits_i{O_i}$~ (disjoint union).
\subsection{Discrete Dynamical Models with Space}
In applications the set of states $\wS$ usually has a special structure of a set of functions
on some space.The following constructions form the basis for all types of 
dynamical systems we consider in this chapter:
\begin{enumerate}
  \item
\emph{Space} is a discrete (basically finite) set of points 
$\X=\set{\x_1, \x_2, \ldots, \x_\XN}$
provided with the structure of an abstract regular ($k$-valent) graph.
	\item 
\emph{Space symmetry group}~ $\sG=\set{\sg_1=\id, \sg_2, \ldots, \sg_\sGN}$ 
is the graph automorphism group: $\sG=\mathrm{Aut}(\X)\leq\Perm{\X}$. 
We assume that $\sG$ acts \emph{transitively} on $\X$.
	\item 
\emph{Local space symmetry group} is defined as the \emph{stabilizer}
 of a vertex $\x_i$ in the space group $\sG$: 
 $g\in\sGloc=\mathrm{Stab}_{\sG}\left(\x\right)$ means $\x_ig=\x_i$.%
\footnote{\label{ontheright}We write group actions \emph{on the right}. 
This, more intuitive, convention is adopted in both \emph{GAP} and \emph{Magma} 
-- the
most widespread computer algebra systems with advanced facilities 
for computational group theory.}
\label{ontherightpage} 
Due to the transitivity all such subgroups are isomorphic and we shall 
denote the isomorphism class by $\sGloc$. This is subgroup of  the 
space symmetry group: $\sGloc\leq\sG$. 
	\item 
Points $x\in\X$ take values in a finite set 
$\lS=\set{\ls_1, \ls_2, \ldots, \ls_\lSN}$ of \emph{local states}.
	\item
\emph{Internal symmetry group} $\iG=\set{\ig_1=\id, \ig_2, \ldots, \ig_\iGN}$  
is a group
	$\iG\leq\mathrm{Sym}\left(\lS\right)$ acting on the set of local states $\lS$. 
	\item 
\emph{States of the whole system} are functions $\sigma(x)\in\lSX=\wS$, 
and the set of evolutions takes the form $
\Evoset=\left(\lSX\right)^\Time=\lS^{\X\times\Time}.$
	\item 
We define the \emph{whole symmetry} groups $\W$ 
unifying space $\sG$
	 and internal $\iG$	symmetries  as equivalence 
	 classes of split group extensions of the form 
	 	$$\id\rightarrow\iGX\rightarrow\W\rightarrow\sG\rightarrow\id,$$
where $\iGX$	is the set of $\iG$-valued functions on $\X$.
(More detailed description of this construction see in Sect. \ref{unigroup})	
\end{enumerate} 
The separation of the set $\wS$ into  ``space'' and ``local states'' is not fundamental
--- it is model- and interpretation-dependent.  An example of a system with a somewhat non-standard
notion of space is a quantum computer. Here the space $\X$ is the set of $\XN$ qubits, the set
of local states $\lS$ is $\set{0,1}$. The whole set of states $\wS=\set{0,1}^\X$ contains 
$2^\XN$ elements.
\subsubsection{Example of Discrete Model with Emergent Space-time.}
Modern fundamental theories, in particular the string theory, provide evidence that 
space is an emergent phenomenon \cite{Seiberg}, arising from more basic concepts. 
We demonstrate here that if we have a concept of time then discrete space-time structures 
may arise under very simple and general assumptions. 
It is sufficient to have a time-labelled sequence of events and ability to distinguish 
different types of the events. 
Then space dimensions arise as the counters of events of different types.
\par
\begin{figure}[!h]
\begin{center}
	\includegraphics[width=0.45\textwidth]{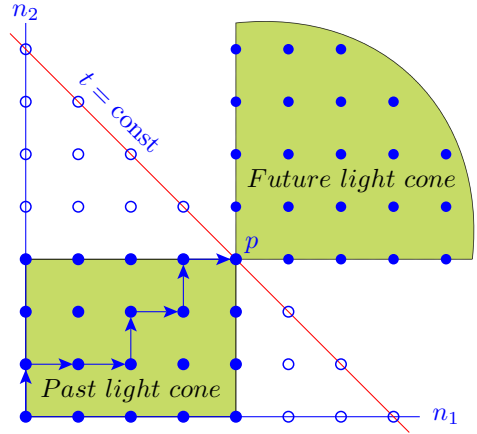}
\caption{Space-time point $p$ is equivalence class of paths with equal numbers of 
$n_1, n_2, \ldots, n_{N+1}$\label{lightcones}.}
\end{center}
\end{figure}

Let us consider a set of states (symbols)
$\Sigma=\set{\sigma_1,\sigma_2,\ldots,\sigma_{N+1}}$ and assume that it is possible 
to observe the sequences (histories) $h=s_0,s_1\cdots{}s_t$, where $s_i\in\Sigma$. 
Let us define a space-time point $p$ as equivalence class of sequences with equal 
numbers of occurrences of each symbol, i.e., $p$ is a ``commutative monomial'' of the 
total degree $t$ described by $N+1$ non-negative integers:
$p=\vect{n_1,\ldots,n_{N+1}},~n_1+\cdots+n_{N+1}=t,$ 
$n_i\in\Z_{\geq0}$ is multiplicity 	of symbol $\sigma_i$ in the history $h$.
The concepts of ``causality'' and ``light cones'' arises naturally. 
The ``speed of light limitation'' is simply impossibility to get more than $t$
symbols (``perceptions'') in $t$ observations --- in terms of monomials the ``past light cone'' 
is the set of divisors of the monomial $p$, the ``future light cone'' is the set of its multiples,
see Fig. \ref{lightcones}.
\par
The union of all possible histories form a \emph{causal network}. As to modelling continuous
Euclidean spaces by this structure, the system of discrete points can be embedded into 
a continuum in many different ways: as a set of discrete points into a
continuous space of arbitrary non-zero dimension%
\footnote{For example, the map $p\rightarrow\alpha_1n_1+\cdots+\alpha_{N+1}n_{N+1}\in\R^1$, 
where $\alpha_i$ are independent irracionals, provides one-to-one embedding of the set of 
points  into $\R^1$.},
as a network into a three-dimensional space%
\footnote{A network, as a locally finite 1-dimensional simplicial complex, can always be embedded into $\R^3$.}.
To separate space from the space-time one should introduce a rule identifying points at 
different times. The identification can be any causality-respecting projection onto
the set $t=\mathrm{const}$. To construct an illustrative discrete model of this section 
we use the following projection. Let us identify the symbols $\sigma_i$ with $N+1$ unit vectors forming regular
simplex in an $N$-dimensional Euclidean space. These systems of vectors (network generating sets)
look like~~ 
\raisebox{0.005\textwidth}{\includegraphics[width=0.07\textwidth]{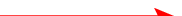}},~~
\raisebox{-0.02\textwidth}{\includegraphics[width=0.07\textwidth]{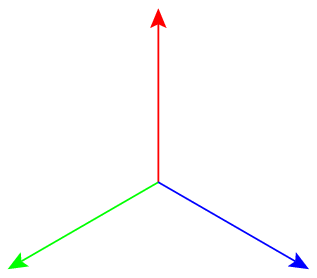}},~~
\raisebox{-0.027\textwidth}{\includegraphics[width=0.07\textwidth]{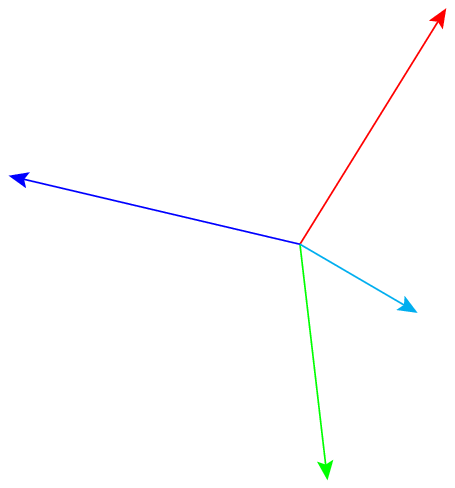}}~~
for $N=1,2,3,$ respectively. The space lattices generated by these sets in four time steps 
$\left(t=4\right)$ for 
the cases $N=2$ and $N=3$
 are shown in the figure 
\begin{center}
\raisebox{0.06\textwidth}{\includegraphics[width=0.3\textwidth]{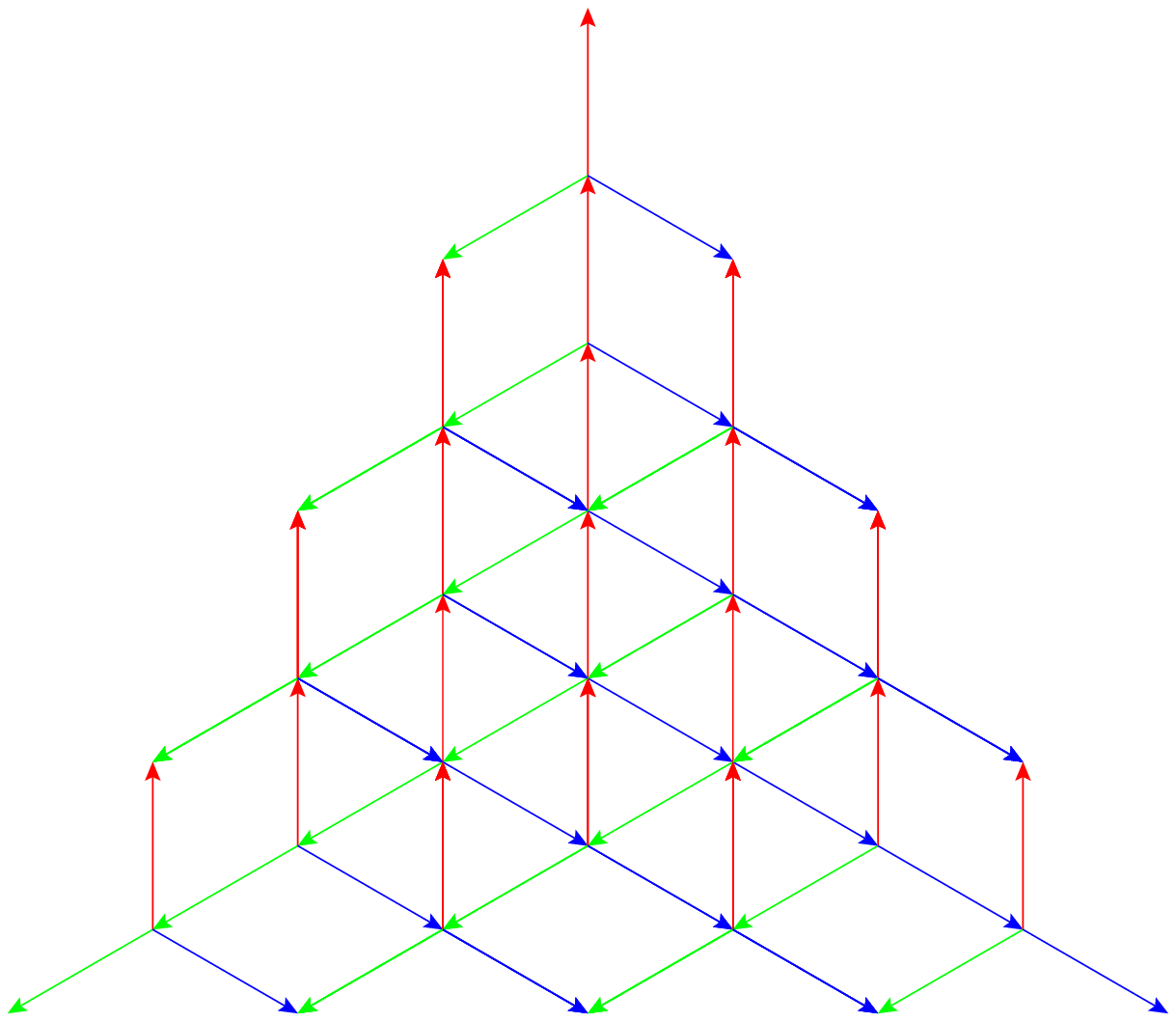}}
\hspace*{0.1\textwidth}
\includegraphics[width=0.3\textwidth]{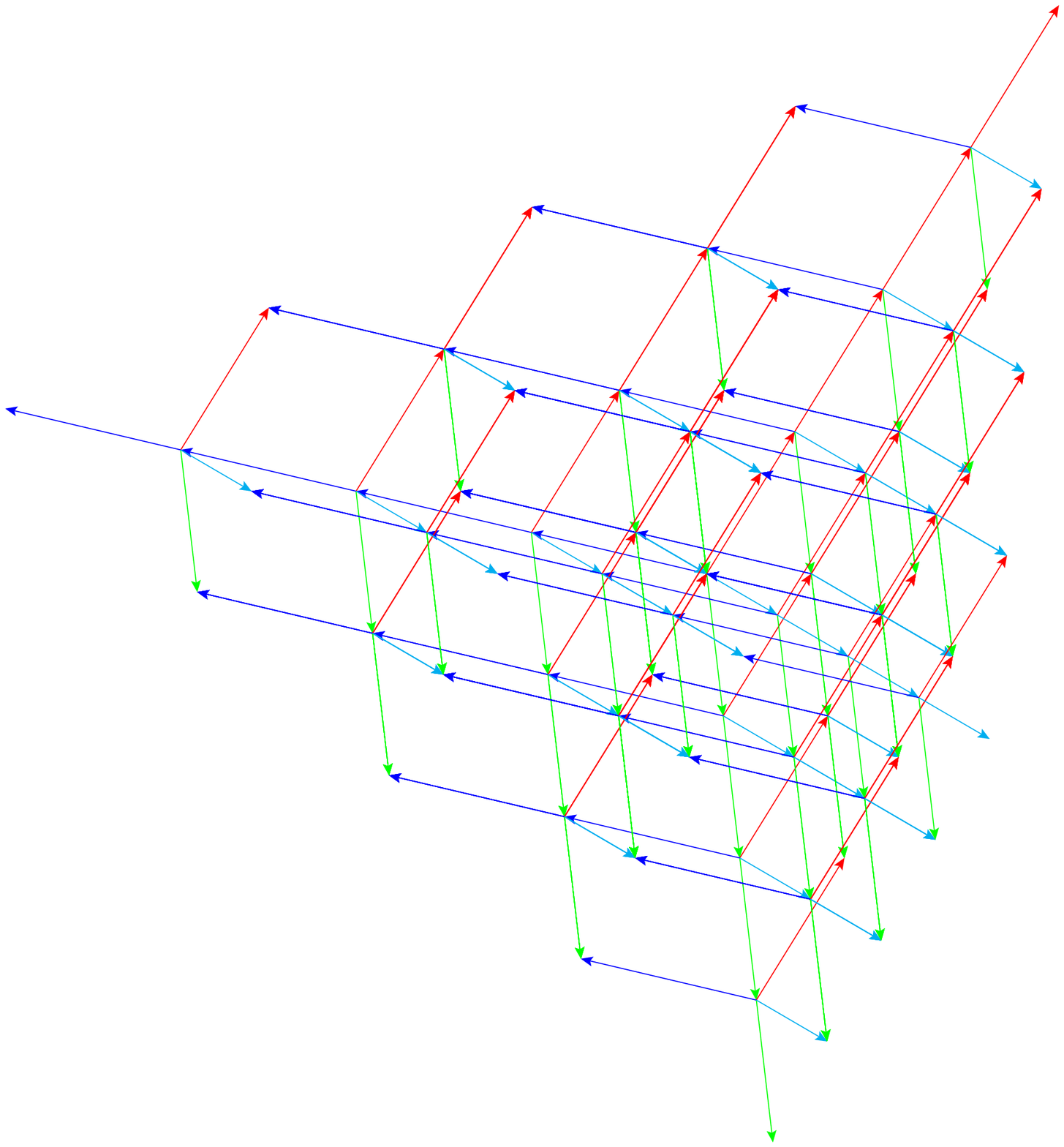}
\end{center}
\par
With these prerequisites, let us construct a simple physical model 
in 1-dimensional space ($N=1$).
We have $\Sigma =\set{\sigma_1,\sigma_2}=\set{{\color{red}\rightarrow},{\color{green}\leftarrow}}$,
~$t=n_1+n_2\in\Z_{\geq0}$. Let us add a little physics by imposing the structure of 
\emph{Bernoulli trials} on the sequences $h=s_0s_1\cdots{}s_t$. Namely, let us introduce
probabilities
$p_1$ and $p_2$ ($\lprob+\rprob=1$) for possible outcomes $\sigma_1$ and $\sigma_2$ of a single trial.
The probability of a separate history $h$ is described by the \emph{binomial distribution}
\begin{equation}
P\left(n_\rstep,n_\lstep\right)=\frac{\left(n_\rstep+n_\lstep\right)!}{n_\rstep!n_\lstep!}
\rprob^{n_\rstep}\lprob^{n_\lstep}.\label{bindistr}
\end{equation}
From this model we can see that the behavior of a discrete system may differ essentially 
from the behavior of its continuous approximation.
Applying~ Stirling's approximation~ to \eqref{bindistr} and introducing new variables 
$x = n_\rstep-n_\lstep,~ \veloc=\rprob-\lprob$
--- let us call them ``space'' and ``velocity'', respectively ---  we obtain
\begin{equation}
	P\left(x,t\right)\approx \PP\left(x,t\right)
	=\frac{1}{\sqrt{1-\veloc^2}}\sqrt{\frac{2}{\pi{}t}}
	\exp\left\{-\frac{1}{2t}\left(\frac{x-\veloc t}{\sqrt{1-\veloc^2}}\right)^2\right\}.
	\label{Pxt}
\end{equation}
This is the \emph{fundamental solution} of the \emph{heat} (also known 
as  \emph{diffusion} or \emph{Fokker--Planck})%
\footnote{The name of the equation depends on interpretation of the function $\PP\left(x,t\right)$.} 
\emph{equation}:
\begin{equation}
	\frac{\partial \PP\left(x,t\right)}{\partial t}
	+\veloc\frac{\partial \PP\left(x,t\right)}{\partial x} 
	=\frac{\left(1-v^2\right)}{2}
	\frac{\partial^2 \PP\left(x,t\right)}{\partial x^2}.
	\label{heq}
\end{equation}
Note that expression \eqref{Pxt} ---  due to 
 the velocity limits $-1\le\veloc\le1$ in our model
 --- contains ``relativistic'' fragment 
$\frac{\textstyle{x-\veloc t}}{\textstyle{\sqrt{1-\veloc^2}}}$. 
Note also that at $\left|\veloc\right|=1$ equation \eqref{heq}
is reduced to the \emph{wave equation}
\begin{equation}
	\frac{\partial \PP\left(x,t\right)}{\partial t}\pm\frac{\partial 
	\PP\left(x,t\right)}{\partial x} 
	=0.
	\label{weq}
\end{equation}
\par
Now let us set a problem as is typical in mechanics: find extremal trajectories connecting 
two fixed points $(0,0)$ and $(X,T)$. As a version of the ``least action principle'', we adopt 
here the search of trajectories of maximum probability. 
The probability of trajectory connecting the points $(0,0)$ and $(X,T)$ and passing through 
some intermediate point $(x,t)$ is the following \emph{conditional probability}
\begin{eqnarray}
			P_{(0,0)\rightarrow(x,t)\rightarrow(X,T)}&=&\frac{P(x,t)P(X-x,T-t)}{P(X,T)}\nonumber\\
				&=&\frac{t!(T-t)!\left(\frac{T-X}{2}\right)!\left(\frac{T+X}{2}\right)!}
			{\left(\frac{t-x}{2}\right)!\left(\frac{t+x}{2}\right)!
			\left(\frac{T-t}{2}-\frac{X-x}{2}\right)!\left(\frac{T-t}{2}+\frac{X-x}{2}\right)!T!}.
			\label{exactp}
\end{eqnarray}
The conditional probability computed for approximation \eqref{Pxt} takes the form
\begin{equation}
	\PP_{(0,0)\rightarrow(x,t)\rightarrow(X,T)}=\frac{T}{\sqrt{\frac{\pi}{2}(1-\veloc^2)tT(T-t)}}
	\exp\left\{-\frac{\left(Xt-xT\right)^2}{2(1-\veloc^2)tT(T-t)}\right\}.
	\label{approxp}
\end{equation}
One can see essential differences between \eqref{exactp} and \eqref{approxp}:
\begin{itemize}
	\item exact probabilities \eqref{exactp} do \emph{not depend} on the velocity $\veloc$
	(or, equivalently,  on the probabilities $\rprob,~\lprob$ of a single trial), whereas \eqref{approxp}
	contains \emph{artificial dependence}, 
	\item it is easy to check that expression \eqref{exactp} allows 
	\emph{many} trajectories
	 with the \emph{same maximum probability}, whereas extremals of \eqref{approxp} are
	  \emph{deterministic trajectories}, namely, straight lines ~$x = \frac{X}{T}t.$
	  This is a typical example of emergence of deterministic behaviour as a result of 
	  the law of large numbers approximation.
\end{itemize}
\subsubsection{Space Symmetries in More Detail.}
\label{SymLat}
A space  $\X$ in our models has the structure of a graph. 
Graphs --- we shall call them also 
\emph{lattices} --- 
are sufficient for all our purposes. In particular, they are adequate to 
introduce  gauge and quantum structures. The symmetry group of the 
space $\X$ is the graph automorphism group 
$\sG=\mathrm{Aut}\left(\X\right)$. The \emph{{automorphism group}} of a 
graph with $n$ vertices may have up to $n!$ elements. Nevertheless, 
the most efficient currently algorithm designed by B. McKay 
\cite{McKay} determines the graph automorphisms by constructing 
compact set (no more than $n-1$ elements, but usually much less)
of generators of the group.
\par
\begin{figure}[!h]
\centering
	\includegraphics[width=0.8\textwidth]{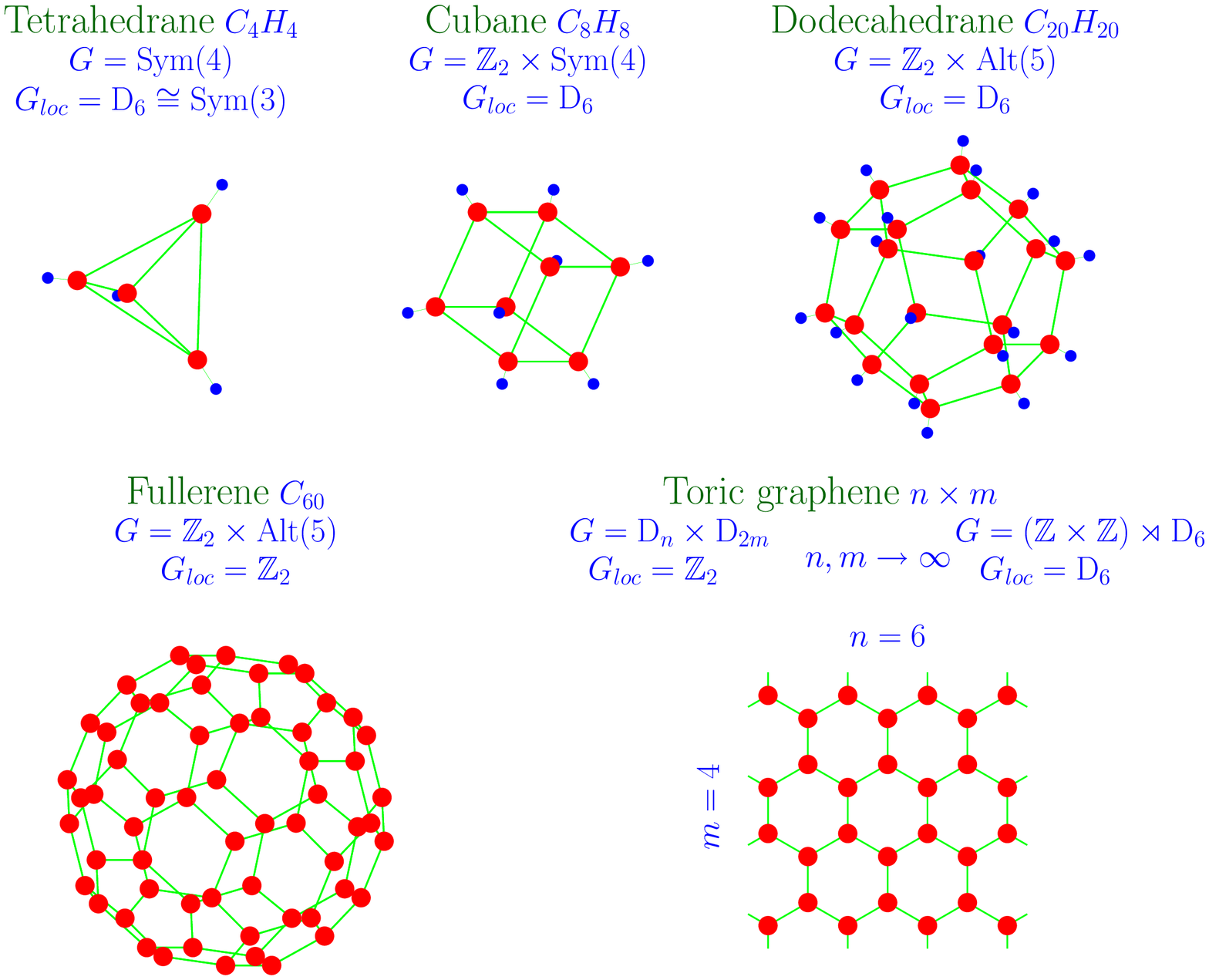}
\caption{Symmetries of 3-valent (hydro)carbon nanostructures.}
	\label{Nanocarbons-1}
\end{figure}

Very often dynamics of a model is expressed in terms of rules defined
on the neighborhoods of lattice vertices.
For this sort of  models with locally defined evolution rules --- 
typical examples are cellular automata and the Ising model --- 
the above mentioned group of local symmetries 
$\sGloc$ is essential. 
Local rules are defined on \emph{orbits} of  $\sGloc$ on 
\emph{edges} from the \emph{neighborhoods} of points  $x.$
Fig. \ref{Nanocarbons-1} shows the symmetry groups $\sG$ and  $\sGloc\leq\sG$ 
for some carbon and hydrocarbon molecules.
\par
Let us consider the role of the local group $\sGloc$ in more detail 
using the \emph{buckyball}\label{buckyball} 
\raisebox{-0.025\textwidth}{\includegraphics[width=0.065\textwidth]{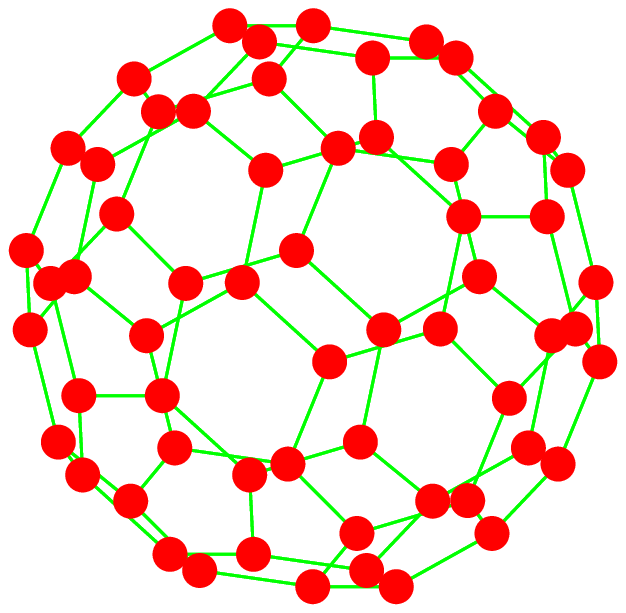}} 
as an example.
The incarnations of this 3-valent graph include in particular:
\begin{itemize}
\vspace*{-10pt}
	\item the \emph{Caley graph of the icosahedral group}%
\footnote{The classical book by F. Klein \cite{Klein} 
is devoted entirely to this group.} 
$\Alt{5}$
(in mathematics); 
	\item  the \emph{molecule of fullerene $C_{60}$} (in carbon chemistry).
\end{itemize}
The symmetry group of the buckyball is $\sG=\mathrm{Aut}\left(\X\right)=\Cgr{2}\times\Alt{5}$.
The neighborhood of a vertex $x_i$ takes the form 
\raisebox{-0.07\textwidth}
{\includegraphics[width=0.15\textwidth]
{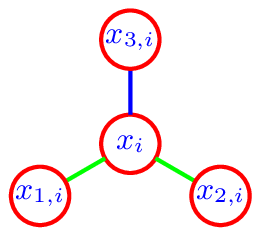}}.
The stabilizer of $x_i$ is $\sGloc=\mathrm{Stab}_{\sG}\left(x_i\right)=\Cgr{2}$. 
The set of neighborhood edges contains three elements: 
$$E_i=\set{e_{1,i} = \inrb{x_i,x_{1,i}},
e_{2,i} = \inrb{x_i,x_{2,i}}, e_{3,i} = \inrb{x_i,x_{3,i}}}.$$
The set of orbits of $\sGloc$ on $E_i$ consists of two orbits:
 $$\Omega_i=\set{\omega_{1,i}
=\set{e_{1,i}, e_{2,i}}, \omega_{2,i}=\set{e_{3,i}}},$$ i.e.,
the stabilizer does not move the edge 
 $\rbra{x_i,x_{3,i}}$ and swaps $\rbra{x_i,x_{1,i}}$ and
 $\rbra{x_i,x_{2,i}}.$ This asymmetry results from different roles the edges
 play in the structure of the buckyball:  $\rbra{x_i,x_{1,i}}$ and
 $\rbra{x_i,x_{2,i}}$ are edges of a pentagon adjacent to
 $x_i$ , whereas $\rbra{x_i,x_{3,i}}$
 separates two hexagons; in the carbon molecule $C_{60}$ the edge $\rbra{x_i,x_{3,i}}$ 
 corresponds to the double bond, whereas others are the single bonds.
\par
Naturally formulated local rules determining behavior of a system must
respect decompositions of  neighborhoods into the orbits of the group of local symmetries.
For example, the Hamiltonian of the Ising model on the buckyball must depend on two, 
generally different, coupling constants $J_{12}$ and
$J_3$. Moreover, the coupling constants may be of different types ---
ferromagnetic or antiferromagnetic --- and this may lead to interesting behavior of the model.  
Such natural Hamiltonian should take the form
\begin{equation}
	H_{bucky} = -\frac{1}{2}\sum\limits_is_i
	\left[J_{12}\left(s_{1,i}+s_{2,i}\right)+J_3s_{3,i}\right] 
	-B\sum\limits_is_i~,\label{Hbucky}
\end{equation}
where $s_i, s_{1,i}, s_{2,i}, s_{3,i}\in\lS=\set{-1,1}.$
In a similar way the local rule for a cellular automaton on the buckyball
must have the form
$$
x_i' = f\left(x_i,x_{1,i},x_{2,i},x_{3,i}\right)~,
$$
where function $f$ must be symmetric with respect to variables
 $x_{1,i}$ and $x_{2,i}$, i.e., $$f\left(x_i,{\color{red}x_{1,i}},{\color{blue}x_{2,i}},x_{3,i}\right)
 \equiv{}f\left(x_i,{\color{blue}x_{2,i}},{\color{red}x_{1,i}},x_{3,i}\right).$$
\subsubsection{Unification of Space and Internal Symmetries.}
\label{unigroup}
Having the groups $\sG$ and $\iG$ acting on $\X$ and $\lS$, 
respectively, we can combine them into a single group $\W\le\Perm{\lSX}$ 
which acts on the states $\wS=\lSX$ of the whole system. The group $\W$ can be 
identified, as a set, with the \emph{Cartesian product} $\iGX\otimes\sG$, 
where $\iGX$ is the set of $\iG$-valued functions on $\X.$ That is, every 
element $u\in\W$ can be represented in the form 
$u = \welem{\alpha(x)}{a},$ where $\alpha(x)\in\iGX$ and $a\in\sG.$ 
\emph{A priori} there are different possible ways to combine $\sG$ and $\iG$ 
into a single group. So selection of possible combinations should be 
guided by some natural (physical) reasons. General arguments 
convince that the required  combination $\W$ should be a \emph{split 
extension} of the group  $\sG$ by  the group $\iGX.$
In physics, it is usually assumed that the space and internal symmetries 
are independent, i.e., $\W$ is the \emph{direct product} $\iG^\X\Times\sG$ 
with action on $\lSX$ and multiplication rules:
\begin{eqnarray}
\sigma(x)\welem{\alpha\left(x\right)}{a}
&=&
\sigma\left(x\right)\alpha\left(x\right)
\text{~~~~~~~~~~~\emph{action}},\nonumber \\
\welem{\alpha\left(x\right)}{a}\wmult\welem{\beta\left(x\right)}{b}
&=&
\welem{\alpha\left(x\right)\beta\left(x\right)}{ab}\label{wmultdir}
\text{~~~~\emph{multiplication}}. 
\end{eqnarray}
Another standard construction is the \emph{wreath product} $\iG\wr_{\X}\sG$ 
having a structure of the semidirect product $\iGX\Rtimes\sG$ with 
action and multiplication
\begin{eqnarray}
\sigma(x)\welem{\alpha\left(x\right)}{a}&=&\sigma
\left(xa^{-1}\right)\alpha\left(xa^{-1}\right),
\nonumber \\
\welem{\alpha\left(x\right)}{a}\wmult\welem{\beta\left(x\right)}{b}&=&
\welem{\alpha\left(x\right)\beta\left(xa\right)}{ab}\label{wmultwr}. 
\end{eqnarray}
These examples are generalized by the following\\
\textbf{Statement.}
\emph{ There are equivalence classes of {split group extensions}
\begin{equation}
	1\rightarrow\iGX\rightarrow\W\rightarrow\sG\rightarrow1
	\label{ext}
\end{equation}
determined by {antihomomorphisms}%
\footnote{The term `antihomomorphism'
means that $\mu(a)\mu(b)=\mu(ba).$}
 $\mu: \sG \rightarrow \sG$.
The equivalence is described by {arbitrary} function $\kappa: \sG \rightarrow \sG.$
The explicit formulas for main group  operations --- {\emph{action}} on $\lSX$, 
{\emph{multiplication}}
and  {\emph{inversion}} --- are
\begin{eqnarray}
	\sigma(x)\welem{\alpha\left(x\right)}{a}&
	=&\sigma\left(x\mu(a)\right)\alpha\left(x\kappa(a)\right),
	\label{wact}\\[3pt]
	\welem{\alpha\left(x\right)}{a}\wmult\welem{\beta\left(x\right)}{b}&=&
\welem{\alpha\left(x\kappa(ab)^{-1}\mu(b)\kappa(a)\right)\beta\left(x\kappa(ab)^{-1}
\kappa(b)\right)}
	{ab}, \label{wmult}\\
	\welem{\alpha(x)}{a}^{-1}
	&=&\welem{\alpha\left(x\kappa\left(a^{-1}\right)^{-1}\mu(a)^{-1}\kappa(a)\right)^{-1}}{a^{-1}}.
	\label{winv}
\end{eqnarray}
}
This statement follows from the general description of the structure of split 
extensions 
of a group $\sG$ by a group $H$:
all such extensions are determined by the homomorphisms from $\sG$ to $\mathrm{Aut}\left(H\right)$ 
(see, e.g., \cite{Kirillov}). Specializing this description to the case when $H$ is 
the set of $\iG$-valued function on $\X$ and $\sG$ acts on arguments  of these functions we obtain our statement.
The \emph{equivalence} of extensions with the same antihomomorfism $\mu$ but with different functions $\kappa$
is expressed by the commutative diagram
\begin{equation}
\begin{diagram}
\node{\id}
\arrow[1]{e}
\node{\iGX}
\arrow[1]{e}
\arrow[1]{s,=}
\node{\W}
\arrow[1]{e}
\arrow[1]{s,l}{K}
\node{\sG}
\arrow[1]{e}
\arrow[1]{s,=}
\node{\id}\\
\node{\id}
\arrow[1]{e}
\node{\iGX}
\arrow[1]{e}
\node{~\W'}
\arrow[1]{e}
\node{\sG}
\arrow[1]{e}
\node{\id}
\end{diagram},
\end{equation}
where the mapping $K$ takes the form
$
K: \welem{\alpha(x)}{a}\mapsto\welem{\alpha\left(x\kappa(a)\right)}{a}.
$
\par
Note that the standard \emph{direct}  and \emph{wreath}  
products are obtained from this general construction by choosing
antihomomorphisms 
$\mu(a)=1$ and $\mu(a)=a^{-1},$ 
respectively. As to the \emph{arbitrary} function $\kappa,$ the choices $\kappa(a)=1$
and $\kappa(a)=a^{-1},$ respectively, 
are generally used in the literature.
\par
In our computer programs (written in C) the group $\W$ is specified by two groups $\sG$ and 
$\iG$ and 
two functions $\mu(a)$ and $\kappa(a)$ implemented as arrays. It is convenient in computations 
 to use the
following specialization: $\mu(a)=a^{-m}$ and $\kappa(a)=a^k$.
For such a choice formulas \eqref{wact}-\ref{winv} take the form 
\begin{eqnarray}
	\sigma(x)\welem{\alpha\left(x\right)}{a}&=&\sigma\left(xa^{-m}\right)\alpha\left(xa^k\right),
	\label{wactspec}\\[3pt]
	\welem{\alpha\left(x\right)}{a}\wmult\welem{\beta\left(x\right)}{b}&=&
	\welem{\alpha\left(x(ab)^{-k-m}a^{k+m}\right)\beta\left(x(ab)^{-k}b^k\right)}
	{ab}, \label{wmultspec}\\
	\welem{\alpha(x)}{a}^{-1}
		&=&\welem{\alpha\left(xa^{2k+m}\right)^{-1}}{a^{-1}}.
	\label{winvspec}
\end{eqnarray}
Here $k$ is \emph{arbitrary} integer, but $m$ is restricted only to two values:
$m=0$  and $m=1$, i.e., 
such specialization does not 
cover other than, respectively, \emph{direct} and \emph{wreath} types of split extentions.
On the other hand, the antihomomorphisms $\mu(a)=1$ and $\mu(a)=a^{-1}$ exist for any
group, while others depend on the particular structure of a group.
Note that actions of $\sG$ on any function $f(x)$ are called \emph{trivial} 
and \emph{natural} for  $\mu(a)=1$ and $\mu(a)=a^{-1}$, respectively.
\section{Structural Analysis of Discrete Relations}
The methods of compatibility analysis, such as the Gr\"obner basis
computation or reduction to involutive form, are widely used to study
systems of polynomial and differential equations.
In this section we develop similar techniques for discrete systems, in particular,
for cellular automata.
\par
Let us consider the Cartesian product $\Sigma^n=\Sigma_1\times{}\Sigma_2\times\cdots\times{}\Sigma_n,$
i.e., the set of ordered $n$-tuples $\left(\sigma_1,\sigma_2,\ldots,\sigma_n\right),$ 
with $\sigma_i\in{}\Sigma_i$
for each $i$. By definition, \emph{$n$-ary relation} is any subset of the $n$-dimensional
hyperparallelepiped $\Sigma^n.$ We assume that $\Sigma_i$ are finite sets of 
$q_i = \left|\Sigma_i\right|$ elements that we shall call \emph{states}. 
\par
We can treat $n$ dimensions of the hyperparallelepiped $\Sigma^n$ as elements of a set of points
$X=\left\{x_1,x_2,\ldots,x_n\right\}$. To make this initially amorphous set into a ``space''
(or ``space-time'')
we should provide $X$ with a structure determining how ``close'' to each other are 
different points.  The relevant mathematical abstraction of such a structure is 
an abstract simplicial complex. 
The natural concept of space assumes the homogeneity of its points. This means that 
there  exists a symmetry group acting transitively on $X$, i.e., providing possibility to ``move''
any point into any other. The homogeneity is possible only if all $\Sigma_i$
are equivalent. Let us denote the equivalence class by $\Sigma$. We can represent 
$\Sigma$ canonically in the form
$\Sigma=\left\{0,\ldots,q-1\right\},~q=\left|\Sigma\right|$.
\par
If the number of states is a prime power,
$q=p^m$, we can additionally equip the set $\Sigma$ with the structure of the Galois
field $\F_q$. Using the functional completeness of polynomials --- this means that 
\emph{any} function can be represented as polynomial --- over finite fields \cite{Lidl},
we can represent any $k$-ary relation on $\Sigma$ as a set of zeros of some polynomial belonging
to the ring $\F_q\left[x_1,\ldots,x_k\right]$. Thus, the set of relations can be regarded as a
system of polynomial equations. Although this description is not necessary (and does not work,
if $\Sigma_i$ are different sets or $q$ is not prime power), it is useful due to our habit to
employ polynomials wherever possible and capability of applying different advanced tools 
of polynomial algebra, such as, for example, the Gr\"obner bases.
\par
An \emph{abstract simplicial complex} (see, e.g., \cite{Hilton}) $K=\left(X,\Delta\right)$
is determined by a set of points $X=\left\{x_1,x_2,\ldots,x_n\right\}$ and an assembly $\Delta$
of subsets of $X$, which are called \emph{simplices}, such that (a) for all $x_i\in{}X$ 
$\left\{x_i\right\}\in\Delta$
and (b) if $\tau\subseteq\delta\in\Delta,$ then $\tau\in\Delta.$ The subsets of a simplex 
--- they are also simplices due to (b) --- are called \emph{faces}. 
Condition (a) means that all one-element subsets are simplices. 
Clearly, the structure
of the complex $K$, i.e., the set $\Delta$, is uniquely determined by the simplices that are
\emph{maximal} by inclusion. \emph{Dimension} of a simplex $\delta$ is the number 
$\dim\delta = \left|\delta\right|-1$. This definition is motivated by the fact that $k+1$
points immersed in the general position into the Euclidean space $\R^{n\geq{}k}$ form a 
$k$-dimensional convex polyhedron. The dimension of a complex $K$ is defined as the maximum
dimension of all simplices in $K$: $\dim{}K=\max\limits_{\delta\in\Delta}\dim\delta$.
From the point of view of abstract combinatorial topology, no matter how the
complex can be immersed into the space $\R^{n}$ --- it is essential only how its simplices are
connected with each other. However, it follows from the N\"{o}beling--Pontryagin theorem 
that any (locally finite) abstract $k$-dimensional
complex can be geometrically realized in the space $\R^{2k+1}$. We will show below that,
for any $n$-ary relation $R\subseteq{}\Sigma^n$, one can regularly and uniquely construct some
abstract simplicial complex.
\subsection{Basic Definitions and Constructions}
In addition to $k$-simplices, which are singled out sets of $k+1$ points, we need to consider
arbitrary sets of point. For brevity, we shall call sets containing $k$ points by $k$-sets.
Dealing with systems of relations defined on different sets of points, it is necessary to
establish correspondence between the points and dimensions of the hypercube $\Sigma^k$. 
This is achieved by using exponential notation.
The notation $\Sigma^{\set{x_i}}$ fixes $\Sigma$ as the set of values of the point $x_i$.
For the $k$-set $\delta=\set{x_1,\ldots,x_k}$, we introduce the notation 
$\Sigma^\delta=\Sigma^{\set{x_1}}\times\cdots\times{}\Sigma^{\set{x_k}}$. 
The set $\delta$ is called the \emph{domain} of the relation
$R^\delta$. We will call the whole hypercube $\Sigma^\delta$
a \emph{trivial} relation. Accordingly, $R^\delta\subseteq{}\Sigma^\delta$ 
denotes a relation given
on the set of points $\delta$.
\subsubsection{Relations.}
Thus, we have:
\par
\textbf{Definition 1} (relation). A \emph{relation} $R^\delta$ on the set of
points $\delta=\set{x_1,\ldots,x_k}$ is any subset of the hypercube $\Sigma^\delta$;
i.e., $R^\delta\subseteq{}\Sigma^\delta$.\\
The relation $R^\delta$ can be regarded as the Boolean-valued function 
$R^\delta: \Sigma^\delta\rightarrow\set{0,1}$.
We can think of $x_i$'s as variables taking values in $\Sigma$
and write the relation as 
$$a=R^\delta\left(x_1,\ldots,x_k\right), ~~a\in\set{0,1}.$$
\par
An important special case of relations:
\par
\textbf{Definition 2} (functional relation).
 A \emph{relation} $R^\delta$ on the set of
points\\ $\delta=\set{x_1,\ldots,x_k}$ is called \emph{functional} if there is a position
$i\in\left(1,\ldots,k\right)$~ such~ that\\ for~ any 
 $\sigma_1,\ldots,\sigma_{i-1},\sigma_{i+1},\ldots,\sigma_{k},\varsigma,\tau\in\Sigma$\\ 
 from~ 
 $\inrb{\sigma_1,\ldots,\sigma_{i-1},\varsigma,\sigma_{i+1},\ldots,\sigma_{k}}\in{}R^\delta$
 and
 $\inrb{\sigma_1,\ldots,\sigma_{i-1},\tau,\sigma_{i+1},\ldots,\sigma_{k}}\in{}R^\delta$\\
it follows~ that $\varsigma=\tau$.\\
In terms of variables the functional relation $R^\delta$ can be written in the form 
$$x_i=F\inrb{x_1,\ldots,x_{i-1},x_{i+1},\ldots,x_{k}}, \text{~~where~~} 
F: \Sigma^{\delta\setminus\set{x_i}}\rightarrow\Sigma.$$
\par
We need to be able to extend relations from subsets
of points to larger sets:
\par
\textbf{Definition 3} (extension of relation). For given set of
points $\delta$, its subset $\tau\subseteq\delta$ and relation $R^\tau$ on the subset $\tau$,
we define the \emph{extension} of $R^\tau$ as the relation
$$R^\delta=R^\tau\times{}\Sigma^{\delta\setminus\tau}.$$
This definition, in particular, allows the relations
$R^{\delta_1},\ldots,R^{\delta_m}$ defined on different domains to be extended
to the common domain, i.e., to the union ${\delta_1}\cup\cdots\cup{\delta_m}$.
\par
Logical implications of the relations are defined in a
natural way:
\par
\textbf{Definition 4} (consequence of relation). A relation $Q^\delta$ is
called a \emph{consequence} of the relation $R^\delta$ if
$R^\delta\subseteq{}Q^\delta\subseteq{}\Sigma^\delta$;
i.e., $Q^\delta$ is arbitrary \emph{superset} of the set $R^\delta$.
\par
The relation  $R^\delta$ may have many different consequences: their
total number (including $R^\delta$ itself and the trivial relation $\Sigma^\delta$) is
evidently equal to $2^{\left|\Sigma^\delta\right|-\left|R^\delta\right|}$.
\par
It is natural to single out the
consequences that can be reduced to relations on
smaller sets of points:
\par
\textbf{Definition 5} (proper consequence). A \emph{nontrivial}
relation $Q^\tau$ is called the \emph{proper consequence} of the relation
$R^\delta$ if $\tau$ is a \emph{proper} subset of $\delta$ (i.e., $\tau\subset\delta$) 
and the relation $Q^\tau\times{}\Sigma^{\delta\setminus\tau}$ is a consequence of $R^\delta$.
\par
We call relations that have no proper consequences the \emph{prime relations}.
\subsubsection{Compatibility of Systems of Relations.}
The compatibility of a system of relations 
can naturally be defined by the intersection of their
extensions to the common domain:
\par
\textbf{Definition 6} (base relation). The \emph{base relation} of the
system of relations\\ $R^{\delta_1},\ldots,R^{\delta_m}$ is the relation
$$
R^\delta=\bigcap\limits_{i=1}^m{}R^{\delta_i}\times{}\Sigma^{\delta_i\setminus\delta},
~~\text{where}~~ \delta=\bigcup_{i=1}^m\delta_i.
$$
Let us make two comments for the polynomial
case $q = p^n$, where the standard tool for the compatibility analysis is
the Gr\"obner basis method:
\begin{itemize}
	\item The compatibility condition determined by the
\emph{base relation} can be represented by a \emph{single} polynomial,
unlike the Gr\"obner basis, which is normally a system
of polynomials.
	\item Any possible Gr\"obner basis of polynomials representing
the relations $R^{\delta_1},\ldots,R^{\delta_m}$ corresponds to some
combination of consequences of the \emph{base relation}.
\end{itemize}
\subsubsection{Decomposition of Relations.}
If a relation has proper consequences, we can try to
express it as far as possible in terms of these consequences, i.e., 
relations on smaller sets of points. To this end we introduce
\par
\textbf{Definition 7} (canonical decomposition). The \emph{canonical
decomposition} of a relation $R^\delta$ with proper consequences
$Q^{\delta_1},\ldots,Q^{\delta_m}$ is the relation
\begin{equation}
	R^\delta=\PF^\delta\bigcap
	\left(\bigcap\limits_{i=1}^m{}Q^{\delta_i}\times\Sigma^{\delta\setminus\delta_i}\right),
	\label{cade}
\end{equation}
where the factor $\PF^\delta$ is defined by the following
\par
\textbf{Definition 8} (principal factor). The \emph{principal factor}
of the relation $R^\delta$ with proper consequences $Q^{\delta_1},\ldots,Q^{\delta_m}$
is the relation
$$
\PF^\delta=R^\delta\bigcup
\left(\Sigma^\delta\setminus\bigcap\limits_{i=1}^m{}Q^{\delta_i}
\times\Sigma^{\delta\setminus\delta_i}\right).
$$
The principal factor is the maximally ``free'' --- i.e., the
closest to the trivial --- relation that, together with the
proper consequences, makes it possible to recover the
initial relation.
\par
If the principal factor in the canonical decomposition
is trivial, the relation is completely reduced to relations
on smaller sets of points.
\par
\textbf{Definition 9} (reducible relation). A relation $R^\delta$ is
said to be \emph{reducible} if it can be represented as
\begin{equation}
R^\delta=
	\bigcap\limits_{i=1}^m{}Q^{\delta_i}\times\Sigma^{\delta\setminus\delta_i},
	\label{rere}	
\end{equation}
where $\delta_i$ are \emph{proper subsets} of $\delta$.
\par
This definition makes it possible to impose a ``topology''
--- i.e., the structure of an abstract simplicial complex
with the corresponding theories of homologies, cohomologies,
etc. --- on an \emph{arbitrary} $n$-ary relation $R\subseteq\Sigma^n$.
This is
achieved by 
\begin{itemize}
	\item naming the dimensions of the hypercube $\Sigma^n$ 
as the ``points'' $x_1,\ldots,x_n\in{}X$,
	\item decomposing $R$ (which can now be denoted by $R^X$) into
\emph{irreducible} components, 
	\item and defining the \emph{maximal
simplices} of the set $\Delta$ as the \emph{domains} of irreducible components
of the relation $R^X$.
\end{itemize}
\subsubsection{On Representation of Relations in Computer.}
A few words are needed about computer implementation
of relations. To specify a $k$-ary relation $R^k$ we should
mark its points within the $k$-dimensional hypercube 
(or hyperparallelepiped) $\Sigma^k$,
i.e., define a \emph{characteristic function}
$\chi: \Sigma^k\rightarrow \set{0,1},$ with $\chi(\vec{\sigma})=1$ or 0
according as $\vec{\sigma}\in R^k$ or 
$\vec{\sigma}\notin R^k$.
Here $\vec{\sigma} = \left(\sigma_0,\sigma_1,\ldots,\sigma_{k-1}\right)$
is a point of the hypercube. The simplest way to implement
the characteristic function is to enumerate all
the $q^k$ hypercube points in some standard,
e.g., lexicographic order. Then the relation can be represented by a string of $q^k$ bits
$\alpha_0\alpha_1\cdots\alpha_{q^k-1}$ in accordance with the table:
\begin{center}
\begin{tabular}{ccccc|c|c}
$\sigma_0$&$\sigma_1$&$\ldots$&$\sigma_{k-2}$&$\sigma_{k-1}$&$i_{\vec{\sigma}}$
&$\chi(\vec{\sigma})$
\\
\hline
$0$&$0$&$\ldots$&
$0$&$0$& 0&$\alpha_0$
\\
$1$&$0$&$\ldots$&
$0$&$0$& 1&$\alpha_1$
\\
$\ \vdots$&$\ \vdots$&$\cdots$&$\ \vdots$&
$\ \vdots$&$\vdots$&$\vdots$
\\
$q-2$&$~q-1$&$\ldots$&
$q-1$&$~q-1$&
$~q^k-2$&$\alpha_{q^k-2}$
\\
$q-1$&$~q-1$&$\ldots$&
$q-1$&$~q-1$&$~q^k-1$&$\alpha_{q^k-1}$
\\
\end{tabular}
\end{center}
 We
call this string \emph{bit table} of relation. Symbolically
$\mathrm{BitTable}\left[\,i_{\vec{\sigma}}\right]:= \left(\vec{\sigma}\in
R^k\right).$ Note that $\vec{\sigma}$ is the (``little-endian'')
representation of the number $i_{\vec{\sigma}}$ in the radix $q$:
$$
i_{\vec{\sigma}}=	\sigma_0+\sigma_1q+\cdots+\sigma_iq^i+\cdots+\sigma_{k-1}q^{k-1}.
$$
In the case of hyperparallelepiped 
$\Sigma^k=\Sigma_1\times\Sigma_2\times\cdots\times\Sigma_k$ 
one should use the \emph{multi-radix representation} of integers:
$$
i_{\vec{\sigma}}=	\sigma_0+\sigma_1\times{}q_1+\cdots
+\sigma_i\times{}q_1q_2\cdots{}q_i
+\cdots+\sigma_{k-1}\times{}q_1q_2\cdots{}q_{k-1},
$$
where $0\leq{}\sigma_i<q_{i+1},~~ i\in\left[0,\ldots{},k-1\right]$.
\par
The characteristic function (bit table) can be represented as the \emph{binary} integer
\begin{equation}
\chi=	\alpha_0+\alpha_12+\cdots+\alpha_i2^i+\cdots+\alpha_{q^k-1}2^{q^k-1}.
\label{BT}
\end{equation}
Most
manipulations with relations are reduced to very efficient bitwise
computer commands. Of course, symmetric or sparse (or, vice versa,
dense) relations can be represented in a more economical way, but
these are technical details of implementation.
\subsection{Illustration: Application to Some Cellular Automata}
\subsubsection{J. Conway's Game of Life.}
The ``Life family'' is a set of 2-dimensional, binary (i.e., $\Sigma=\set{0,1}$; 
$q=2$) cellular automata
similar to \CL, which rule is defined on 9-cell (3$\times$3) 
Moore neighborhood and is described as follows. A
cell is ``born'' if it has exactly 3 ``alive'' neighbors, ``survives''
if it has 2 or 3 such neighbors, and ``dies'' otherwise. This rule is
symbolized in terms of the ``birth''/``survival'' lists as
B3/S23. Another examples of automata from this family are
\emph{\textbf{HighLife}} (the rule B36/S23), and
\emph{\textbf{Day\&Night}} (the rule B3678/S34678).
 The site \cite{lives-site} contains collection of more than 
 twenty rules from the Life family with Java applet to run these rules and
 descriptions of their behavior.
\par
Generalizing this type of local rules, we define a \emph{$k$-valent Life
rule} as a \emph{binary} rule on a $k$-valent neighborhood (we adopt that 
$x_1,\ldots,x_{k}$ are neighbors of $x_{k+1}$ of the central cell $x_{k+1}$) 
described by two \emph{arbitrary} subsets of the set $\set{0,1,\ldots,k}$. 
These subsets $B,S\subseteq\set{0,1,\ldots,k}$ contain conditions 
for the one-time-step 
transitions $x_{k+1}\rightarrow{}x^\prime_{k+1}$ 
of the forms $0\rightarrow1$ and
$1\rightarrow1$, respectively. Since the number of subsets of any finite
set $A$ is $2^{\left|A\right|}$ and \emph{different} pairs $B$/$S$ 
define \emph{different} rules, the number of different rules defined by two sets
$B$ and $S$ is equal to $2^{k+1}\times2^{k+1}$.  
Thus, the total number of $k$-valent rules described
by the ``birth''/``survival'' lists is
\begin{equation}
N_{B/S,\>k}= 2^{2k+2}. \label{NBS}
\end{equation}
\par
There is another way to characterize this type of local rules. 
Let us consider $k$-valent 
rules symmetric with respect to the group $\Perm{k}$ of all permutations 
of $k$ outer points of the neighborhood. We shall call such rules $k$-symmetric. 
It is not difficult to count
the total number of different $q$-ary $k$-symmetric rules:
\begin{equation}
N_{q,\>\Perm{k}}=q^{\binom{k+q-1}{q-1}q}.
		\label{nqk}	
\end{equation}
We see that \eqref{nqk} evaluated at $q=2$ coincides with \eqref{NBS}, i.e., 
$N_{2,\>\Perm{k}}=N_{B/S,\>k}$.
Since $k$-valent Life rules are obviously $k$-symmetric we have the following\\
\textbf{Proposition.} \emph{For any $k$ the set of $k$-symmetric binary rules coincides
with the set of $k$-valent Life rules.}
\par
This proposition implies in particular that one can always express any
 $k$-symmetric binary rule in terms of the ``birth''/``survival'' lists.
\par
The local relation of \CL~automaton $R^\delta_{CL}$
is defined on the 10-set $\delta=\set{x_1,\ldots,x_{10}}$:
\begin{center}
\includegraphics[width=0.37\textwidth]{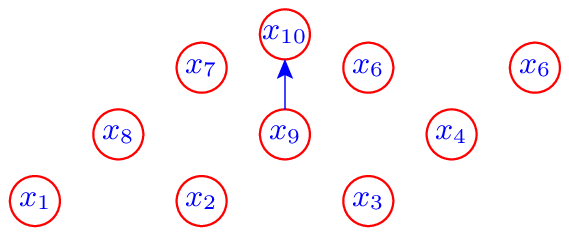}
\end{center}
Here the point $x_{10}\equiv{}x^\prime_9$ is the next-time-step
of the point $x_9$. By construction, elements of the 10-dimensional 
hypercube $\Sigma^\delta$ belong to the relation of \CL~automaton, 
i.e.,  $\inrb{x_1,\ldots,x_{10}}\in{}R^\delta_{CL}$, in the following cases: 
\begin{enumerate}
    \item
$\left(\sum_{i=1}^8x_i = 3\right) \wedge \left(x_{10} = 1\right)$,
    \item
$\left(\sum_{i=1}^8x_i = 2\right) \wedge \left(x_9 = x_{10}\right)$,
    \item
$x_{10} = 0$, if none of the above conditions holds.
\end{enumerate}
The number of elements of the relation $\RCL$ is  $\left|\RCL\right|=512$.
The relation $\RCL$, as is the case
for any cellular automaton, is \emph{functional}:
the state of $x_{10}$ is uniquely determined by the states of other points.
The state set $\Sigma=\set{0,1}$ can be \emph{additionally}
endowed with the structure of the field $\F_2.$
We accompany the below analysis of the structure of
$\RCL$ by description in terms of
polynomials from
 $\F_2\left[x_1,\ldots,x_{10}\right].$
This is done only for illustrative purposes and for comparison
with the Gr\"obner basis method.
In fact, we transform the
relations into polynomials only for output. 
Transformation of any relation into polynomial form
can be performed by computationally very cheap multivariate version 
of the Lagrange interpolation. In the case $q=2$, the polynomial which set
of zeros corresponds to a relation is constructed uniquely. If
$q=p^n>2$, there is a freedom in the choice of nonzero values of
constructed polynomial, and the same relation can be represented
by many polynomials.
\par
The polynomial representing $\RCL$ takes the form
\begin{equation}
\PCL= x_{10}
+x_9
\left(
\esp_7
+\esp_6
+\esp_3
+\esp_2
\right)
+\esp_7
+\esp_3,
\label{polylife}
\end{equation}
where $\esp_k\equiv\esymm{k}{x_1,\ldots,x_8}$ is
the $k$th \emph{elementary symmetric polynomial} defined
for  $n$ variables
$x_1,\ldots,x_{n}$ by the formula:
$$
\esymm{k}{x_1,\ldots,x_{n}} =
\sum\limits_{1\leq{}i_1<i_2<\cdots<i_{k}\leq{}n}x_{i_1}x_{i_2}\cdots x_{i_{k}}.
$$
Hereafter, we will use the following notation:
$$
\esp_k\equiv\esymm{k}{x_1,\ldots,x_{8}},~~~~
\esp_k^i\equiv\esymm{k}{x_1,\ldots,\widehat{x_i},\ldots,x_{8}},
$$
$$
\esp_k^{ij}\equiv\esymm{k}{x_1,\ldots,\widehat{x_i},\ldots,\widehat{x_j},\ldots,x_{8}}. 
$$ 
Applying the computer program to $\RCL$, we find that the relation $\RCL$ is 
\emph{reducible} and
has the decomposition
\begin{equation}
\RCL =
R_2^{\delta\setminus\set{x_9}}\bigcap
\left(\bigcap\limits_{k=1}^7
R_1^{\delta\setminus\set{x_{i_k}}}\right),
\label{lifedecomp}
\end{equation}
where $\inrb{i_1,\ldots,i_7}$ is arbitrary 7-element subset of the set $\inrb{1,\ldots,8}$.
For brevity, we dropped in \eqref{lifedecomp} the trivial factors $\Sigma^{\set{x_{i_k}}}$
entering into the general formula \eqref{rere}.
\par
The eight relations $R_1^{\delta\setminus\set{x_{i}}}$ ($1\leq{}i\leq8$; for decomposition
\eqref{lifedecomp}, it suffices to take any seven of them) have
the following polynomial form:
$$
x_9x_{10}
\left(\esp^i_6+\esp^i_5+\esp^i_2+\esp^i_1\right)
+x_{10}\left(\esp^i_6+\esp^i_2+1\right)
+x_9\left(\esp^i_7+\esp^i_6+\esp^i_3+\esp^i_2\right)=0.
$$
Accordingly, the relation $R_2^{\delta\setminus\set{x_9}}$ has the form
$$
x_{10}\left(\esp_7+\esp_6+\esp_3+\esp_2+1\right)+\esp_7+\esp_3=0.
$$
\par
The relations $R_1^{\delta\setminus\set{x_{i}}}$ and $R_2^{\delta\setminus\set{x_9}}$ 
are \emph{irreducible} but \emph{not prime}, and can be expanded in accordance with
formula \eqref{cade}.
Continuing the decomposition iterations, we finally
obtain the following system of relations (in the polynomial
form) that are satisfied for \CL:
\begin{eqnarray}
x_9x_{10}\left(\esp^i_2+\esp^i_1\right)
+x_{10}\left(\esp^i_2+1\right)
+x_9\left(\esp^i_7+\esp^i_6+\esp^i_3+\esp^i_2\right)&=&0,
\label{poly1red}
\\
x_{10}\left(\esp_3+\esp_2+1\right)+\esp_7+\esp_3&=&0,
\label{poly2red}
\\
\left(x_9x_{10}+x_{10}\right)\left(\esp^{ij}_3+\esp^{ij}_2+\esp^{ij}_1+1\right)&=&0,
\label{poly11red}
\\
x_{10}\left(\esp^{i}_3+\esp^{i}_2+\esp^{i}_1+1\right)&=&0,
\label{poly12red}
\\
x_{10}x_{i_1}x_{i_2}x_{i_3}x_{i_4}&=&0.
\label{poly0123red}
\end{eqnarray}
One can easily interpret the simplest relations \eqref{poly0123red}: 
if the point $x_{10}$ is in the state 1, then at least
one point in any set of four points surrounding $x_9$ must be in the state 0.
\par
The above analysis of the relation $\RCL$
takes $< 1$ sec on a 1.8GHz AMD Athlon notebook with 960Mb.
\par
To compute the Gr\"obner basis we must add to polynomial
\eqref{polylife} ten polynomials
$$
 x_i^2+x_i,\ i = 1,\ldots,10
$$
corresponding to the relation $x^{q}=x$ that holds
for all elements of any finite field $\F_{q}$.
\par
Computation of the Gr\"obner basis over $\F_2$ with the
help of \textbf{Maple 9} gives the following:
\begin{itemize}
	\item pure lexicographic order with variable ordering
$x_{10}\succ x_9\succ\cdots\succ x_1$ does not provide any 
new information leaving initial polynomial
\eqref{polylife} unchanged;
	\item  pure lexicographic order with variable ordering  
	$x_1\succ x_2\succ\cdots\succ x_{10}$ reproduces
relations \eqref{poly1red}---\eqref{poly0123red} (modulo
several polynomial reductions violating the symmetry
of polynomials); the computation takes 1 h 22 min;
	\item degree-reverse-lexicographic
order also reproduces system
\eqref{poly1red}---\eqref{poly0123red} (same comment as above); 
the times are: 51 min
for the variable ordering $x_1\succ x_2\succ\cdots\succ x_{10}$, and 33 min
for the ordering $x_{10}\succ x_9\succ\ldots\succ x_1$.
\end{itemize}
\subsubsection{Elementary Cellular Automata.}
Simplest binary, nearest-neighbor, 1-dimensional
cellular automata were named \emph{elementary cellular automata}
by S. Wolfram, who has extensively studied their properties
\cite{Wolfram}. A large collection of results concerning these automata
is presented in Wolfram's online atlas \cite{site}.
In the exposition below we use Wolfram's notations and terminology.
The elementary cellular automata
are simpler than \CL, and we
may pay more attention to the topological aspects
of our approach.
\par
Local rules of the elementary cellular automata are defined
on the 4-set
$\delta=\set{p,q,r,s}$ which can be pictured by the icon
\raisebox{-0.02\textwidth}{\includegraphics[width=0.1\textwidth]{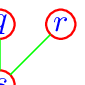}}.
A local rule is a binary function of the form $s=f(p,q,r).$
There are totally $2^{2^3}=256$ such functions,
each of which can be indexed with an 8-bit binary number.
\par
Our computation with relations representing the local
rules shows that the total number 256 of them is divided
into 118 reducible and 138 irreducible relations.
Only two of the irreducible relations appeared to be prime,
namely, the rules 105 and 150
in Wolfram's numeration. This numeration is based on the ``big-endian''
--- i.e., opposite to our convention \eqref{BT} --- 
representation of binary numbers.
Note, that the prime rules 105 and 150 have linear polynomial
forms: $s=p+q+r+1$ and  $s=p+q+r$, respectively.
\par
We consider the elementary automata on a space-time
lattice with integer coordinates
$(x,t)$, i.e., 
$x\in \Z$ 
or
$x\in \Z_m$ (spatial $m$-periodicity), $t\in \Z.$
We denote a state of the
point on the lattice by $u(x,t)\in \Sigma=\set{0,1}$.
Generally the points  are connected
as is shown in the picture
\begin{center}
\includegraphics[width=0.3\textwidth]{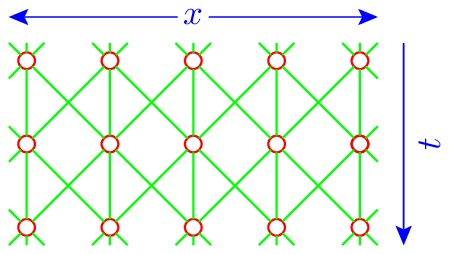}.
\end{center}
The absence of horizontal ties expresses the independence of ``space-like'' 
points in cellular automata.

\paragraph{Reducible Automata.}
The analysis shows that some automata with reducible
local relations can be represented as unions of
automata defined on disconnected subcomplexes:
\par
\begin{itemize}
	\item  Two automata 0 and 255 are determined by
unary relations $s=0$ and $s=1$ on the disconnected set of points:
\begin{center}
\includegraphics[width=0.24\textwidth]{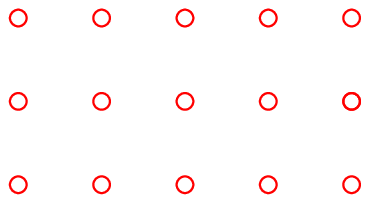}.
\end{center}
Note that  unary relations are usually called \emph{properties}.
\par
	\item  Six automata 15, 51, 85, 170, 204 and 240 are, in fact,
disjoint collections of spacially zero-dimensional automata, i.e., 
single cells evolving in time.
As an example, let us consider the automaton 15.
The local relation is defined on the set
\raisebox{-0.02\textwidth}{\includegraphics[width=0.1\textwidth]{ECASetpqrsColor}} 
and its bit table  is 0101010110101010.
This relation is reduced to the relation on the face
\raisebox{-0.013\textwidth}
{\includegraphics[width=0.062\textwidth]{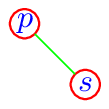}}
with bit table 0110. The space–time
lattice is split in the following way:
\begin{center}
\includegraphics[width=0.24\textwidth]{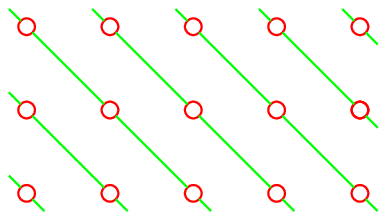}.
\end{center}
The bit table 0110 means that the points $p$ and $s$ can
be only in opposite states, and we can 
write immediately the general solution for the automaton 15:
$$
u(x,t) = a(x-t)+t\mod 2,
$$
where $u(x,0)\equiv{}a(x)$ is an arbitrary initial condition.
\par
	\item  Each of the ten automata 5, 10, 80, 90, 95, 160, 165, 175,
245, 250 is decomposed into two identical  automata.
As an example let us consider the rule 90.
This automaton is distinguished as producing the fractal
of topological dimension 1 and Hausdorff dimension
$\ln3/\ln2\approx1.58$ known
as the \emph{Sierpinski sieve} (or \emph{gasket} or \emph{triangle}).
Its local relation on the set
\raisebox{-0.02\textwidth}{\includegraphics[width=0.1\textwidth]{ECASetpqrsColor}} 
is represented by the bit table 1010010101011010.
The relation is reduced to the relation  
with the bit table
\begin{equation}
10010110 \text{~~~~on the face~~
\raisebox{-0.02\textwidth}{\includegraphics[width=0.1\textwidth]{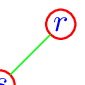}}}.
\label{bt90}
\end{equation}
It can be seen from the structure of face \eqref{bt90} that the
space–time lattice is split into two identical independent
complexes as is shown
\begin{center}
\includegraphics[width=0.8\textwidth]{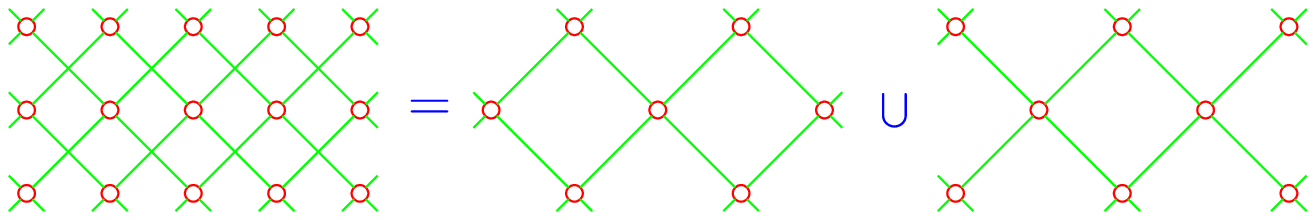}.
\end{center}
To find a general solution of the automaton 90 it
is convenient to use the polynomial form of relation \eqref{bt90} $s+p+r=0$.
With this linear expression, the general
solution is easily constructed:
$$
u(x,t)=\sum\limits_{k=0}^t \binom{t}{k}a(x-t+2k)\mod 2,\qquad{} u(x,0)\equiv{}a(x).
$$
\end{itemize}
\paragraph{Using Proper Consequences.}
Proper consequences --- even if they are not functional --- can provide useful information
on the behavior of a cellular automaton.
\par
For example, 64 automata%
\footnote{The complete list of these automata in Wolfram's numeration
is as follows:
  2,   4,   8,  10,  16,  32,  34,  40,
 42,  48,  64,  72,  76,  80,  96, 112,
128, 130, 132, 136, 138, 140, 144, 160,
162, 168, 171, 174--176, 186, 187,
190--192, 196, 200, 205, 206, 208,
220, 222--224, 234--239,
241--254.}
(with both
reducible and irreducible local relations) have proper consequences
with the bit table
\begin{equation}
1101
\label{finiterod}
\end{equation}
on, at least, one of the faces
\begin{equation}
\text{\includegraphics[width=0.062\textwidth]{ECASetpsColor}
~~~~
\includegraphics[width=0.027\textwidth]{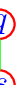}
~~~~
\includegraphics[width=0.062\textwidth]{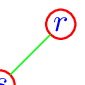}}.
\label{finiterodsets}
\end{equation}
The algebraic forms of relation
\eqref{finiterod} on faces \eqref{finiterodsets} are
$
ps+s=0,~~qs+s=0,~~rs+s=0,
$ respectively.
\par
\begin{figure}[!h]
\begin{center}
\includegraphics[width=0.6\textwidth]{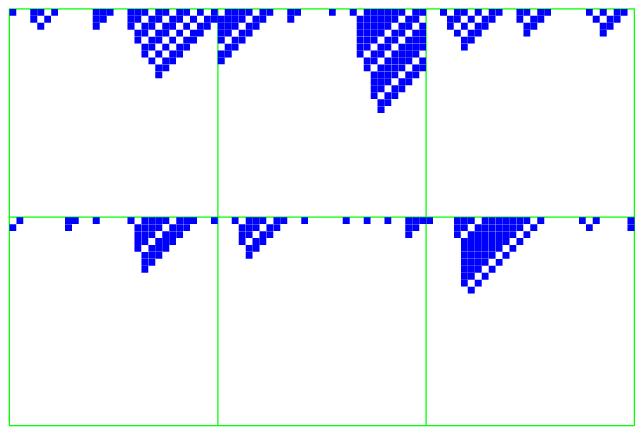}
\caption{Rule 168. Several random initial conditions\label{figu}}
\end{center}
\end{figure}

Relation \eqref{finiterod} is \emph{not functional}, and hence can not 
describe any deterministic evolution.
Nevertheless, it imposes severe restrictions on the behavior
of the automata having such proper consequences.
The features of the behavior resulting from relation \eqref{finiterod} 
are clearly seen from many of computational results presented 
in the atlas \cite{site}.  A typical pattern from this atlas
is reproduced in Fig. \ref{figu}, where several evolutions of the
automaton 168 are presented.
In the figure, 0's and 1's are denoted by the empty and filled square cells, 
respectively. Note that the authors of the figure use
a spatially periodic condition: $x\in\Z_{30}$.
\par
The local relation of the automaton
168 --- its polynomial form is $pqr+qr+pr+s=0$ --- has the proper consequence $rs+s=0$.
Relation \eqref{finiterod} means that if,
say $r$, as for the rule 168, is in the state 1 then $s$ may be
in both states 0 or 1, but if the state of $r$ is 0,
then the state of $s$ must be 0:
$$r=1\Rightarrow{}s=0\vee{}s=1,$$
$$r=0\Rightarrow{}s=0.$$
One can see that all evolutions in Fig. \ref{figu} consist of diagonals $x = x_0-t$ 
directed leftward 
and downward. Each diagonal begins with a several units, but after the first 
appearance of zero all subsequent points along the diagonal are zeros.
\paragraph{Canonical Decomposition vs. Gr\"obner Basis.}
In this paragraph we compare our canonical decomposition 
\eqref{cade} with Gr\"obner basis in the polynomial case. 
Let us begin with two examples of elementary cellular automata.
The Gr\"obner bases are computed in the total degree and reverse lexicographical 
order of monomials. The trivial polynomials $p^2+p,$ $q^2+q,$ $r^2+r$ and $s^2+s$ 
are omitted in the  Gr\"obner bases descriptions.
\begin{itemize}
\item 
\textbf{Automaton 30} is remarkable by its chaotic
behavior and is even used as a random number generator 
in \textbf{\emph{Mathematica}}.
\par
Relation: $1001010101101010$ ~or~  $qr+s+r+q+p=0$.
\par
\textbf{Canonical Decomposition:}
\par
Proper consequences:
\par
\begin{tabular}{lll}
face&
\includegraphics[width=0.062\textwidth]{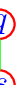}
&
\includegraphics[width=0.1\textwidth]{ECASetprsColor}
\\[10pt]
bit table&11011110 & 11011110\\[10pt]
polynomial \hspace*{20pt}&$qs+pq+q$\hspace*{20pt} & $rs+pr+r$.\\[10pt]
\end{tabular}
\par
Principal factor: $1011111101111111$ ~or~ $qrs+pqr+rs+qs+pr+pq+s+p=0.$
\par
\textbf{Gr\"obner basis:}
$\set{qr+s+r+q+p,~qs+pq+q,~rs+pr+r}.$\\[10pt]
Thus for the rule 30 the polynomials of the canonical decomposition 
coincide (modulo obvious polynomial substitutions) 
with the Gr\"obner basis.
\item 
 \textbf{Automaton 110} is, like a Turing machine, \emph{universal}, i.e., it 
can simulate any computational process, in particular,
any other cellular automaton.
\par
Relation: $1100000100111110$ ~or~ $pqr+qr+s+r+q=0.$
\par
\textbf{Canonical Decomposition:}
\par
Proper consequences:
\par
\begin{tabular}{llll}
face&
\includegraphics[width=0.062\textwidth]{ECASetpqsColor}
&
\includegraphics[width=0.1\textwidth]{ECASetprsColor}
&
\includegraphics[width=0.062\textwidth]{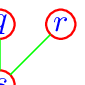}
\\[10pt]
bit table&11011111 & 11011111 &10010111\\[10pt]
polynomial&
$pqs+qs+pq+q$ &
$prs+rs+pr+r$ & $qrs+s+r+q$.\\[10pt]
\end{tabular}
\par
Principal factor:
$1111111111111110$ ~or~ $pqrs=0.$
\par
\textbf{Gr\"obner basis:}
$$\set{prs+rs+pr+r,~qs+rs+r+q,~qr+rs+s+q,~pr+pq+ps}.$$
For automaton 110, the polynomials of the Gr\"obner
basis are not identical with the polynomials of the
canonical decomposition.
The system of relations defined by the
Gr\"obner basis
is:
\begin{eqnarray*}
R_1^{\set{p,r,s}}&=&11011111=\left(prs+rs+pr+r=0\right),
\\
R_2^{\set{q,r,s}}&=&10011111=\left(qs+rs+r+q=0\right),
\\
R_3^{\set{q,r,s}}&=&10110111=\left(qr+rs+s+q=0\right),
\\
R_4^{\set{p,q,r,s}}&=&1110101110111110=\left(pr+pq+ps=0\right).
\end{eqnarray*}
\end{itemize}
In general, the folowing differences between our approach and the Gr\"obner basis 
method can be mentioned.
\begin{itemize}
	\item In contrast to a Gr\"obner basis, 
a base relation, defined as intersection of conditions, 
agrees with the standard in logic and set theory notion of compatibility.
	\item 
In contrast to a canonical decomposition a Gr\"obner basis may 
look beyond the polynomial context as a collection of accidental supersets.
	\item
There is some analogy between  Gr\"obner bases	 and canonical decompositions
 --- in fact, they coincide in about half of cases in our computations. 
	\item
Canonical decomposition is more efficient for problems
with polynomials of arbitrary degree --- the above computation with Conway's
automaton is an example.
  \item
For small degree problems  with large number $n$ of indeterminates the Gr\"obner 
basis outperforms  canonical 
decomposition --- the number of polynomials of bounded degree is a polynomial function 
of $n$, whereas  the algorithm of canonical decomposition scans exponential number
$q^{\textstyle{n}}$ of the hypercube points.
\end{itemize}
 
\section{Soliton-like Structures in Deterministic Dynamics} 
Symmetries of deterministic systems
impose severe restrictions on the system dynamics \cite{Kornyak08}. 
In particular, for the first order%
\footnote{This means that evolution relation \eqref{funcrel} takes the form 
$s_t = F\left(s_{t-1}\right)$.}
functional relations: 
\begin{itemize}
	\item \emph{dynamical trajectories} pass group
orbits in  \emph{non-decreasing} order of orbit sizes,
	\item 	\emph{periodic} trajectories lie within orbits of the \emph{same size}.
\end{itemize}
One of the characteristic features of dynamical systems with non-trivial symmetries
is formation of moving form-preserving structures.
\par
\begin{figure}[!h]
\centering
\includegraphics[width=0.65\textwidth]{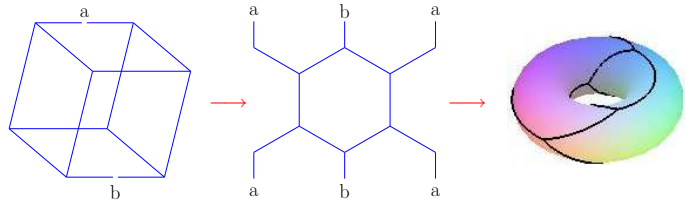} 
\caption{The graph of cube forms 4-gonal (6 tetragons) lattice in sphere
$\S^2$ and 6-gonal (4 hexagons) lattice in torus $\T^2$.}
	\label{Cube-on-tor}
\end{figure}

Let us begin with a simple example. Consider a cube $\X$ whose 
vertices take values in two-element set, say $\lS=\set{0,1}.$
By the way, as is clear from Fig. \ref{Cube-on-tor}, a cube can be interpreted as 
a simplest ``finite model of graphene''.
The 48-element symmetry group of a cube has the structure 
$\sG=\Z_2\times\Perm{4}.$ The group is generated by 3 elements: 
\begin{enumerate}
	\item $120^o$ rotation around diagonal of the cube; 
	\item $90^o$ rotation around axis passing through the 
	centers of opposite cube faces; 
	\item reflection interchanging opposite faces of the cube.
\end{enumerate}
Total number of states of the model is 
$\left|\lSX\right|= 2^8=256.$ If we assume that the group $\iG$ is
trivial, then $\W = \iGX\rtimes\sG=1\rtimes\sG\cong \sG$. 
The group $\W$ splits the set $\lSX$
into 22 orbits in accordance with the table:~~
\begin{tabular}{c|ccccccc}
Size of   orbits&1&2&4&6&8&12&24
\\\hline
Number of orbits&2&1&2&2&5& 4& 6
\end{tabular}~~.
\par
Let us consider a \emph{deterministic} dynamical system on the cube,
namely, symmetric binary 3-valent cellular automaton with the rule 86.
The number 86 is the ``little endian'' representation of the bit string 
01101010 taken from the last column of the rule table with 
$\Perm{3}$-symmetric combinations of values for 
$x_{1,i}, x_{2,i}, x_{3,i}$
\begin{center}
\begin{tabular}[t]{cccc|l}
$x_{1,i}$&$x_{2,i}$&$x_{3,i}$&$x_i$&$x'_i$
\\
\hline
0&0&0&0&$0$\\
0&0&0&1&$1$\\
1&0&0&0&$1$\\
1&0&0&1&$0$\\
1&1&0&0&$1$\\
1&1&0&1&$0$\\
1&1&1&0&$1$\\
1&1&1&1&$0$\\
\end{tabular}
\\[-1pt]\hspace*{130pt}.
\end{center}
Here $x_i$ is value of $i$th vertex of the cube; 
$x_{1,i},x_{2,i},x_{3,i}$ are values of the cube vertices adjacent 
to the $i$th one and $x'_i$ is the next time value of $i$th vertex. 
The rule can also be represented in \CL{~} style 
``Birth''\!/``Survival'' notation as B123/S0,
or as polynomial over the field $\mathbb{F}_2$
$$
x'_i = x_i+\esp_3+\esp_2+\esp_1,
$$
where 
$\esp_1 = x_{1,i}+x_{2,i}+x_{3,i},\ \esp_2 = 
x_{1,i}x_{2,i}+x_{1,i}x_{3,i}+x_{2,i}x_{3,i},\ \esp_3 = x_{1,i}x_{2,i}x_{3,i}$ 
are elementary symmetric functions.
\par
\begin{figure}[!h]
\centering
\includegraphics[width=340pt]{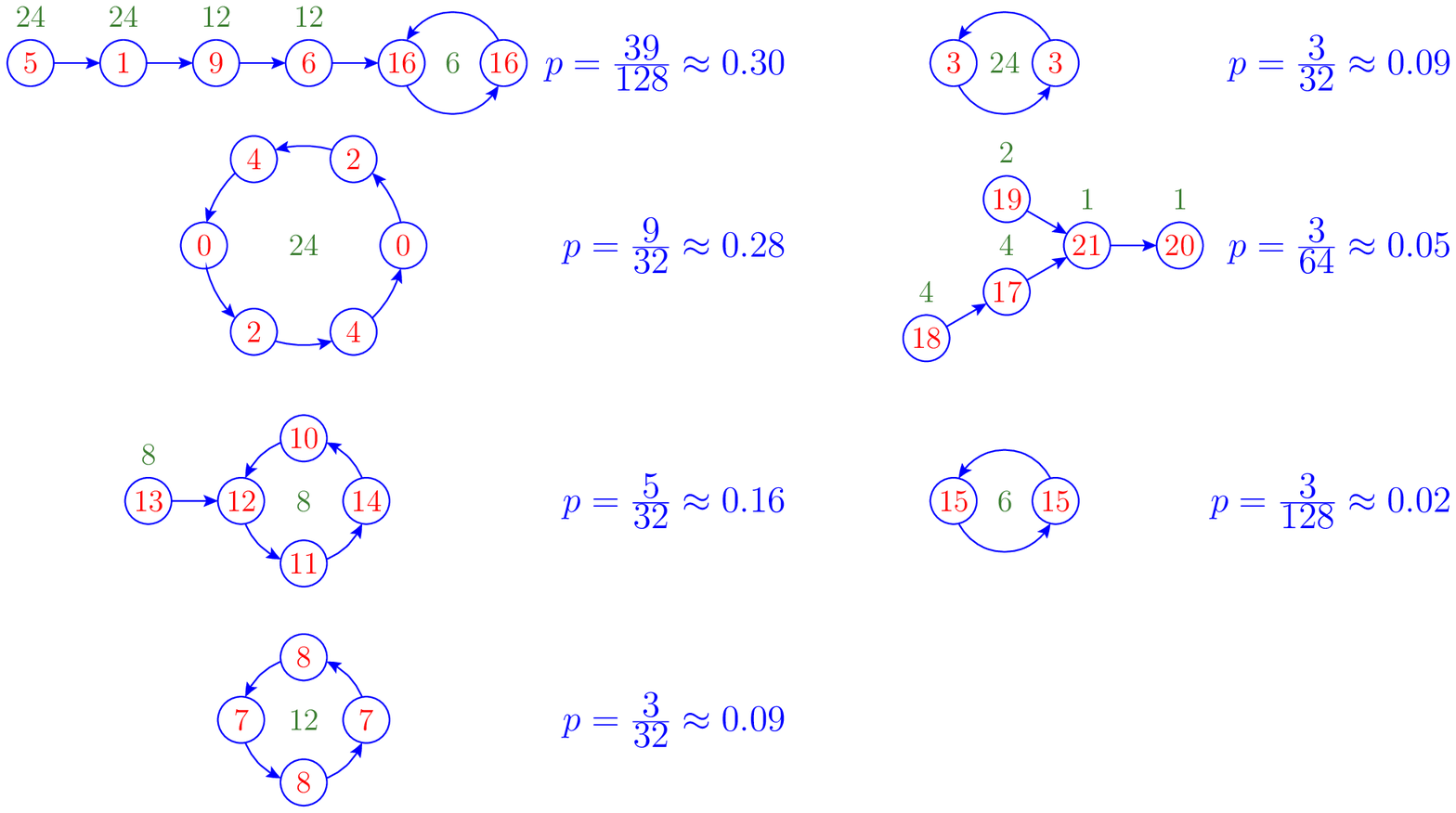}%
\caption{Rule 86. Equivalence classes of trajectories on hexahedron.
}
\label{PhasePortrait}
\end{figure}

The phase portrait of the automaton is shown in Fig. \ref{PhasePortrait}, 
where the group  orbits are represented by circles containing the ordinal numbers%
\footnote{These numbers are specified by 
the computer program in the course of computation.} of orbits within. 
The numbers over orbits and within cycles are sizes of the orbits
(recall that all orbits belonging to the same cycle have equal sizes
--- see the beginning of this section). 
The rational number $p$ indicates the \emph{weight} of the corresponding 
element of the phase portrait. In fact, $p$ is a probability for randomly chosen  
state to appear in an isolated cycle or to be caught by an attractor: 
$p$ = (\emph{size of basin})/(\emph{total number of states}). 
Here \emph{size of basin} is sum of sizes of orbits involved in the struture. 
\par
Generalizing this example,
we see that
if the symmetry group $\W$ splits the 
state set $\lSX$ of \emph{deterministic} dynamical system 
into \emph{finite} number of orbits,  then after some lapse of time \emph{any} trajectory 
comes \emph{inevitably} to a cycle over some finite sequence of orbits. 
This just means formation of \emph{soliton-like structures}.
Namely, let us consider evolution
\begin{equation}
	\sigma_{t_0}(x)\rightarrow\sigma_{t_1}(x)=A_{t_1t_0}\left(\sigma_{t_0}(x)\right).
	\label{evol}
\end{equation}
If the states at the moments $t_0$ and $t_1$ belong to the \emph{same orbit}: 
$\sigma_{t_0}(x)\in{}O_i$ and $\sigma_{t_0}(x)\in{}O_i,$ $O_i\subseteq\lSX;$
then evolution \eqref{evol} can be replaced  by the 
\emph{group action}
$$
\sigma_{t_1}(x)=\sigma_{t_0}(x)w,~~w\in{}W,
$$
i.e., the initial state (``shape'') $\sigma_{t_0}(x)$ is reproduced after some 
``\emph{movement}'' in the space  $\lSX.$
\par
The following are several examples (including continuous cases) of cycles over group orbits: 
\begin{itemize}
	\item  \emph{traveling waves} $\sigma(x-vt)$ in mathematical physics
	---  the Galilei group;
	\item  	``\emph{generalized coherent states}'' in quantum physics
	 --- unitary representations of compact Lie groups;
	\item  		 ``\emph{spaceships}''	in cellular automata  --- lattice symmetries.
\end{itemize}
\par
Let us consider the ``glider'' 
 --- one of the ``spaceships'' 
in \CL{} automaton.
\par 
The space $\X$ of \CL{} is a square lattice. For the finiteness, 
we shall assume that the lattice is closed into the $N\times{}N$ torus.
In the general case $N\neq4$ the symmetry group of $\X$ is the semidirect product of 
two-dimensional translations $\mathrm{T}^2=\Z_N\times\Z_N$ and  the dihedral group 
$\Dih{8} = \Z_4\rtimes\Z_2$: 
\begin{equation}
	\sG=\mathrm{T}^2\rtimes\Dih{8},\mbox{~~if~~} N = 3,5,6,\ldots,\infty.
	\label{genG}
\end{equation}
In the case $N=4$ the translation subgroup $\mathrm{T}^2 = \Z_4\times\Z_4$ is 
\emph{not normal} and $\sG$ has a bit more complicated structure \cite{KornyakCASC09}:
\begin{equation}
	\sG = \overbrace{
	\left(\left(\left(\left(\Z_2 \times \Dih{8}\right) 
	\rtimes \Z_2\right) \rtimes{\color{red}\Z_3}\right) \rtimes \Z_2\right)
	}^{
	\mbox{normal closure
	of~~}\textstyle\mathrm{T}^2}
	\rtimes \Z_2.
	\label{4G}
\end{equation}
The extra symmetry $\Cgr{3}$ in \eqref{4G} can be explained by 
the $\Cgr{3}$ symmetry of the four-vertex Dynkin diagram 
$D_4=~~$\raisebox{-0.025\textwidth}{\includegraphics[width=0.06\textwidth]{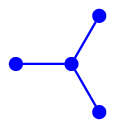}}
 ~~associated with the case $N=4$.
\par
The set of local (cell) states of \CL{} is 
$\lS=\set{\text{``dead''},\text{``alive''}}=\set{0,1}$.
Since the local rule of \CL{} is not symmetric with respect to the transposition
$0\leftrightarrow1$ of the local states, the internal symmetry group is trivial, i.e., 
$\iG=\set{\id}$ and hence $\iGX=\set{\id}$.
Thus, we have  $\W = \iGX\rtimes\sG=\id\rtimes\sG\cong\sG$.
The natural action of $\W$ on functions $\sigma(x)\in\lSX$ takes the form
$\sigma(x)w=\sigma\left(xg^{-1}\right)$, where $w=(\id,g),~g\in\sG$.
\par
\begin{figure}[!h]
\centering
\includegraphics[width=0.7\textwidth]{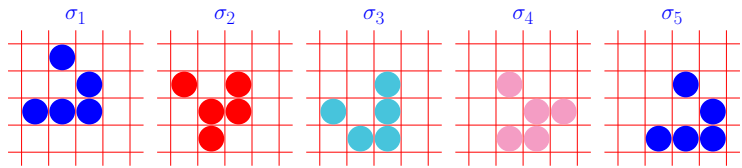}
\caption{Example of soliton-like structure.
``Glider'' in \CL{} is cycle in \emph{two} orbits of the group
 $\sG=\mathrm{T}^2\rtimes\Dih{8}$: configurations $\sigma_3$ and  
$\sigma_4$ are obtained from  $\sigma_1$ and  $\sigma_2$,
respectively, by \emph{the same} combination of downward \emph{shift}, 
$90^o$ clockwise \emph{rotation} and \emph{reflection} in respect to vertical.}
	\label{Glider-2}
\end{figure}

Fig. \ref{Glider-2} shows four steps of evolution of the glider. 
The figure demonstrates how the evolution is reduced to the group action. $N>4$ is
assumed.
\par
\subsubsection*{Comments on Reversibility in Discrete Systems.} 
A typical deterministic dynamical system is \emph{irreversible}
--- it's phase portrait \emph{modulo} group orbits looks like in Fig.
 \ref{PhasePortrait}. We see there several isolated and limit cycles 
 (fixed points are regarded as cycles of unit length) 
 accompanied by influxes flowing into the limit cycles.
In contrast to continuous systems, any discrete system ``forgets'' 
influxes after some time and appears in either isolated or limit cycles. 
After loss of information about  influxes both types of cycles became physically 
indistinguishable and the system behaves just like reversible. This might be a hint 
for explanation of observable reversibility of the fundamental laws of nature.
\par
In this connection we would like to mention recent works of G. 't Hooft.
One of the difficulties of quantum gravity is a conflict between irreversibility of gravity
 --- information loss at the black hole horizon --- with reversibility and unitarity
of the standard quantum mechanics. In several papers of recent years 
(see, e.g., \cite{tHooft99,tHooft06}) 't Hooft developed an approach to 
reconciling both theories. The approach is based on the following assumptions
 \begin{itemize}
	\item physical systems have \emph{discrete degrees of freedom} at tiny (Planck) distance scales;
	\item the states of these degrees of freedom form \emph{primordial} basis
	of Hilbert space (with nonunitary evolution);
	\item primordial states form \emph{equivalence classes}: two states are equivalent if they
	evolve into the same state after some lapse of time;
	\item the equivalence classes by construction form basis of Hilbert space with unitary
	 evolution described by time-reversible Schr\"odinger equation.
\end{itemize}
In our terminology this corresponds to transition to limit cycles: in a
finite time of evolution the limit cycle becomes physically indistinguishable 
from reversible isolated cycle --- the system ``forgets'' its pre-cycle history. 
\par
This type of irreversibility hardly can be observed experimentally (assuming,
of course, that considered models may have at all any relation to physical reality).
The system should probably spend time of order the Planck unit ($\approx10^{-44}$ sec) 
out of a cycle and potentially infinite time on the cycle. Nowadays, the shortest 
experimentally fixed time 
is about $10^{-18}$ sec or $10^{26}$ Planck units.
  
\section{Mesoscopic Lattice Models}
Discrete symmetry analysis simplifies manipulations with 
\emph{microcanonical ensembles} and  search of \emph{phase transitions}.
This allows to reveal subtle details in behavior of \emph{mesoscopic models}.
\label{mesosect}
\subsection{Statistical Mechanics}
As we mentioned earlier, the state of deterministic dynamical system at any point 
of time is determined uniquely by previous states of the system. A Markov chain 
--- for which transition from any state to any other is possible with some probability 
--- is a typical example of \emph{non-deterministic} dynamical system. 
In this section we apply symmetry approach to the lattice models in statistical mechanics. 
These models can be regarded as special instances of Markov chains. 
\emph{Stationary distributions} of the Markov chains are studied by the methods of 
statistical mechanics. 
\par
The main tool of conventional statistical mechanics is the Gibbs \emph{canonical ensemble} 
--- imaginary collection of identical systems placed in a huge thermostat with temperature $T$. 
The statistical properties of canonical ensemble are encoded in the \emph{canonical partition function}
\begin{equation}
Z=\sum\limits_{\sigma\in \lSX}\mathrm{e}^{-E_\sigma/k_B T}.
\label{cpf}
\end{equation} 
Here $\lSX$ is the set of microstates, $E_\sigma$ is energy of microstate $\sigma$, 
$k_B$ is Boltzmann's constant. 
The canonical ensemble is essentially asymptotic concept: its formulation is based on 
approximation called ``thermodynamic limit''. For this reason, the canonical ensemble 
approach is applicable only to large (strictly speaking, infinite) homogeneous systems.
\subsection{Mesoscopy}
Nowadays much attention is paid to study systems which are too large for a 
detailed microscopic description but too small for essential features of 
their behavior to be expressed in terms of classical thermodynamics.
This discipline ---often called \emph{mesoscopy} ---
covers wide range of applications from nuclei, 
atomic clusters and nanotechnological structures to 
multi-star systems \cite{Imry02,Gross01,Gross04}.
To study \emph{mesoscopic} systems one should use 
more fundamental \emph{microcanonical ensemble} instead of canonical one.
A microcanonical ensemble is a collection of identical isolated systems at fixed energy.
Its definition does not include any approximating assumptions. In fact, the only key 
assumption of a microcanonical ensemble is that all its microstates are equally probable. 
This leads to the \emph{entropy} formula
\begin{equation}
S_E=k_{B}\ln\Omega_E,
\label{entropy}
\end{equation}
or, equivalently, to the \emph{microcanonical partition function}
\begin{equation}
\Omega_E=\mathrm{e}^{S_E/k_{B}}.
\label{mcpf}
\end{equation}
Here $\Omega_E$ is the number of microstates at fixed energy $E:$~ 
$\sum\limits_E\Omega_E=\lSXN$. 
In what follows we will omit Boltzmann's constant assuming $k_{B}=1$.
Note that in the thermodynamic limit the microcanonical and canonical
descriptions are equivalent and the link between them is provided by
the Laplace transform. On the other hand, mesoscopic systems demonstrate 
experimentally and computationally observable peculiarities of behavior 
like heat flows from cold to hot, negative specific heat or 
``convex intruders'' in the entropy versus energy diagram,  etc. 
These anomalous --- from the point of view of canonical thermostatistics --- 
features have natural explanation within  microcanonical statistical mechanics \cite{Gross04}.
\subsubsection{Lattice Models.}
In this section we apply symmetry analysis to study mesoscopic lattice
models. Our approach is based on exact enumeration of group orbits of microstates.
Since statistical studies are based essentially on different simplifying assumptions, 
it is important to control these assumptions by exact computation, wherever possible.
Moreover, we might hope to reveal subtle details
in behavior of system under consideration with the help of exact computation. 
\par
As an example, let us consider the Ising model. 
The model consists of \emph{spins} placed on a lattice. 
The set of vertex values is $\lS=\set{-1,1}$ and the interaction 
Hamiltonian
is given by
\begin{equation}
H=-J\sum\limits_{(i,j)}s_is_j - B\sum\limits_{i}s_i,
\label{hamising}
\end{equation} 
where $s_i,s_j\in \lS$; $J$ is a coupling constant ($J > 0$ and $J < 0$ 
correspond to \emph{ferromagnetic} and \emph{antiferromagnetic} cases, respectively);
the first sum runs over all edges $(i,j)$ of the lattice;
$B$ is an external ``magnetic'' field. The second sum $M = \sum\limits_{i}s_i$ 
is called the \emph{magnetization}. To avoid unnecessary technical details we will 
consider only the case $J > 0$
(assuming $J=1$) and $B=0$ in what follows.
\par
Let us  remind that if the local symmetry group $\sGloc$ decomposes 
the sets of edges of lattice neighborhoods into nontrivial orbits, 
then the interaction Hamiltonian should be modified (see, e.g., Eq. 
\eqref{Hbucky} on page \pageref{Hbucky}).
\par
Since Hamiltonian and magnetization are constants on the group
orbits, we can count numbers of microstates corresponding to particular
values of these functions -- and hence compute all needed statistical 
characteristics -- simply by summation of sizes of appropriate orbits.
\par
\begin{figure}[!h]
\centering
\includegraphics[width=0.8\textwidth]{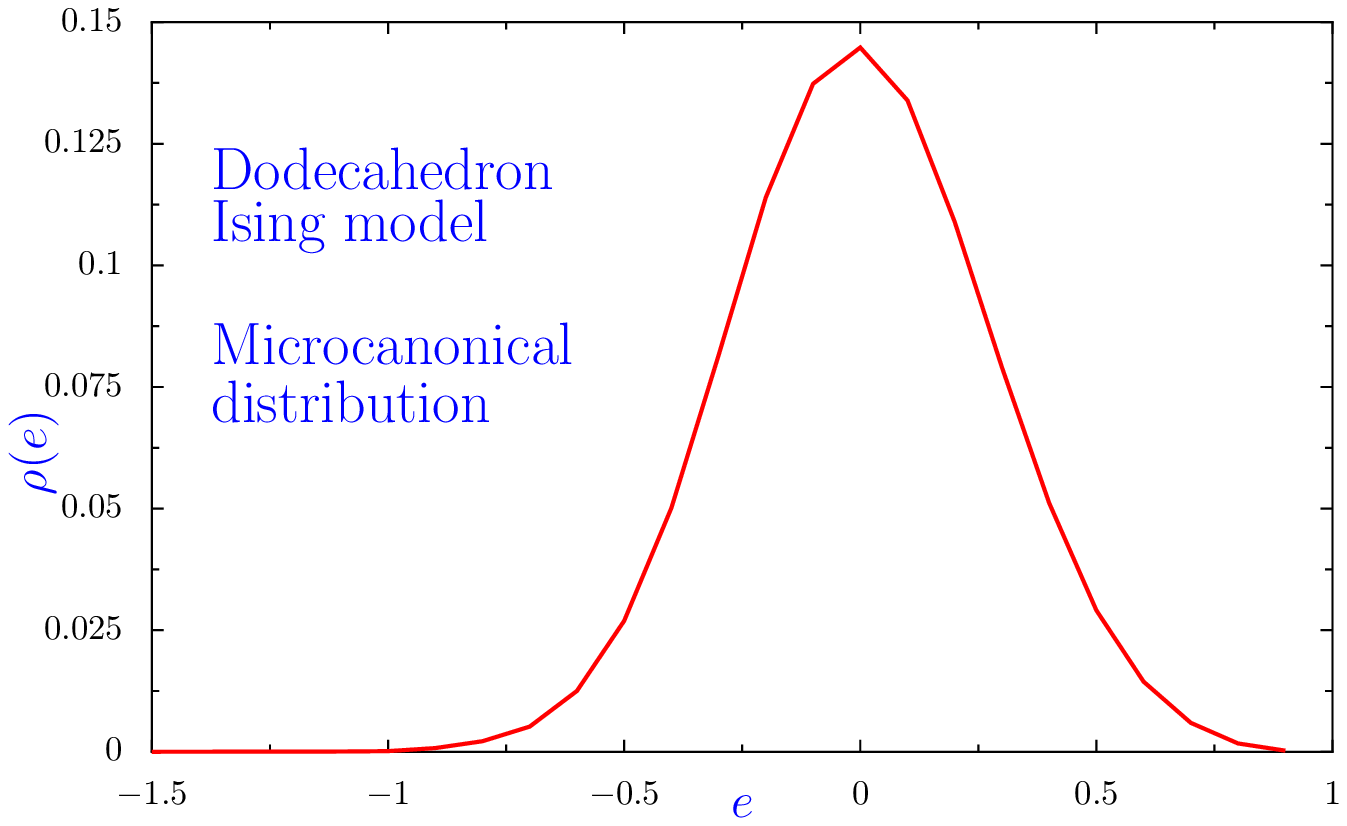}
\caption{Ising model on dodecahedron.
Microcanonical distribution.}
	\label{Microcanonical distribution}
\end{figure}

Fig. \ref{Microcanonical distribution} shows microcanonical partition function
for the Ising model on the dodecahedron
\begin{center}
\includegraphics[width=0.15\textwidth]{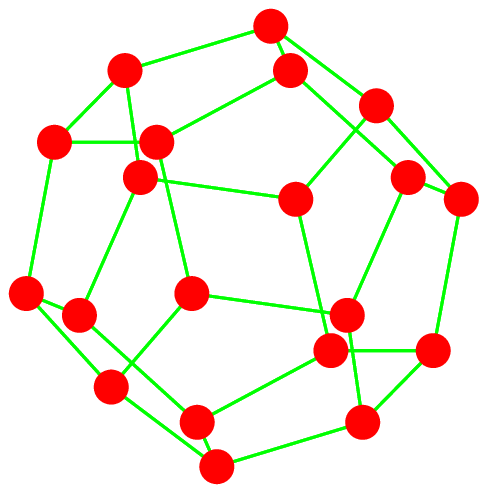}.
\end{center}
 Here total number of microstates
$\lSXN=1048576$, number of lattice vertices $N_\X=20$, 
energy $E$ is value of Hamiltonian.
\par
Of course, other characteristics of the system can be computed easily in this way.
\subsection{Phase Transitions} 
Needs of nanotechnological science 
and nuclear physics attract special
attention to phase transitions in finite systems. Unfortunately
classical thermodynamics and the rigorous theory of critical phenomena 
in homogeneous infinite 
systems fails at the mesoscopic level. Several approaches have been 
proposed to identify phase
transitions in mesoscopic systems. Most accepted of them is search of 
\emph{``convex intruders''} \cite{GrossVotyakov} in
the entropy 
versus energy 
diagram. In the standard thermodynamics there is a relation
\begin{equation}
\left.\frac{\partial^2S}{\partial E^2}\right|_V = -\frac{1}{T^2}\frac{1}{C_V},
\label{d2sde2}
\end{equation}
where $C_V$ is the specific heat at constant volume. 
\par
Relation \eqref{d2sde2} 
implies
that $\left.\partial^2S/\partial E^2\right|_V < 0$ and hence the entropy versus energy 
diagram must be concave. Nevertheless, in mesoscopic systems there might be intervals 
of energy where $\left.\partial^2S/\partial E^2\right|_V > 0$. 
These intervals correspond to first-order 
phase transitions and are called \emph{``convex intruders''}. 
From the point of view of standard thermodynamics one can say about 
phenomenon of \emph{negative heat capacity}, of course, if one accepts that 
it makes sense to define the variables $T$ and $C_V$ as temperature and 
the specific heat at these
circumstances. In \cite{IspolatovCohen} it was demonstrated via computation 
with exactly solvable lattice models that the convex intruders flatten and 
disappear in the models with local interactions as the lattice size grows, 
while in the case of long-range interaction these peculiarities survive
even in the limit of an infinite system (both finite and long-range 
interacting infinite systems are typical cases of systems called 
\emph{nonextensive} in statistical mechanics).
\begin{figure}[!h]
\centering
\includegraphics[width=0.8\textwidth]{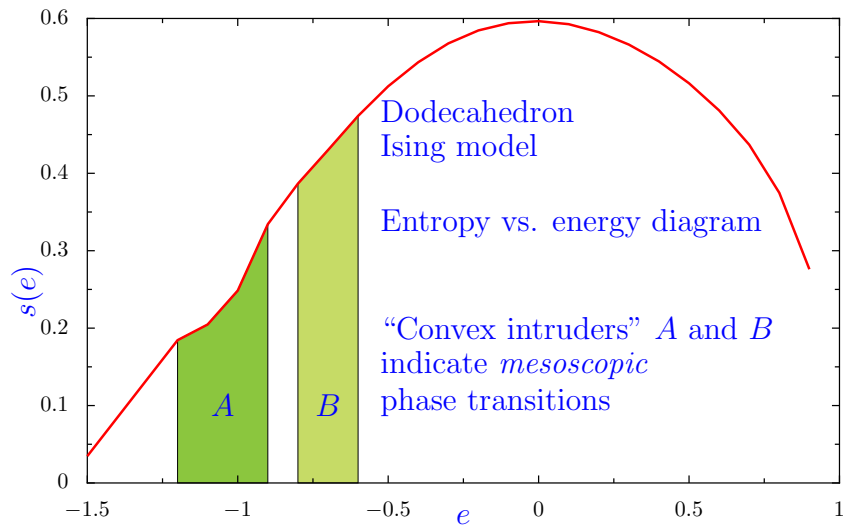}
\caption{Ising model on dodecahedron.
``Convex intruders'' on entropy vs. energy diagram indicate 
\emph{mesoscopic} phase transitions.}
	\label{Convex intruders}
\end{figure}
\par
A convex intruder can be found easily by computer for the discrete systems 
we discuss here. Let us consider three adjacent values of energy 
$E_{i-1}, E_{i}, E_{i+1}$ and corresponding numbers 
of microstates
$\Omega_{E_{i-1}}, \Omega_{E_{i}}, \Omega_{E_{i+1}}$. In our discrete 
case the ratio 
$\frac{E_{i+1}-E_{i}}{E_{i}-E_{i-1}}$ is always rational number $p/q$ 
and 
we can write the convexity condition for entropy in terms of numbers 
of microstates as easily evaluated
inequality
\begin{equation}
\Omega_{E_{i}}^{p+q} < \Omega_{E_{i-1}}^p\Omega_{E_{i+1}}^q.
\label{convcond}
\end{equation}
As a rule $E_{i+1}-E_{i}=E_{i}-E_{i-1}$ and inequality \eqref{convcond} 
takes the form 
$$
\Omega_{E_{i}}^2 < \Omega_{E_{i-1}}\Omega_{E_{i+1}}.
$$
This form means that within convex intruder the number of states with 
the energy $E_{i}$ is less than \emph{geometric mean} of numbers of
states at the neighboring energy levels.
\par
Fig. \ref{Convex intruders} shows the entropy vs. energy diagram 
for the Ising model on dodecahedron. The diagram has apparent convex 
intruder $A$ in the specific energy interval $\left[-1.2,-0.9\right]$.
Exact computation reveals also a subtle convex intruder $B$ in the 
interval $\left[-0.8,-0.6\right]$. 
\section{Gauge Connection and Quantization}
All~ most successful contemporary theories in fundamental physics are gauge theories.
There are also numerous applications of gauge theories in mathematics 
(topological quantum field theory, invariants of 3- and 4-manifolds, monoidal 
categories, Hopf algebras and quantum groups,  etc. \cite{Oeckl}).
\par
In fact, the gauge principle expresses the very general idea that in spite
of the fact that any observable data are represented
in different ``reference frames'' at different points in space%
\footnote{Consideration only time evolution of general 
set of states $\wS$ leads to the trivial
gauge structures. Gauge theories of interest are possible if there exists
underlying space structure, i.e., $\wS=\lSX$.} and time, there 
should be some way to compare these data.
\subsection{Discrete Gauge Principle}
At the set-theoretic level, i.e., in the form suitable for both discrete 
and continuous cases, 
the main concepts of the gauge principle can be reduced to the following. We have
\begin{itemize}
  \item 
a set $\Time$,~ discrete or continuous \emph{time},~ $\Time\cong\Z$ or $\Time\cong\R$;
  \item a set $\X$, \emph{space};
  \item the sets $\Time$ and $\X$ are combined into a space-time $\SpaceTime=\X\times\Time$;
  \item a set $\lS$, \emph{local states};
  \item 
a group $\iG\leq\Perm{\lS}$ acting on $\lS$, \emph{internal symmetries}; 
	\item 
identification of data describing the states from $\lS$ makes 
sense only \emph{modulo} symmetries from $\iG$ --- this is arbitrariness in the choice
of a ``\emph{reference frame}'';
  \item 
there is no \emph{a priori} connection between data (i.e., between reference frames) at different
points $x,~y\in\SpaceTime$ --- we should impose this \emph{connection}
(or \emph{parallel transport}) explicitly as $\iG$-valued function on edges (pairs of points)
of abstract graph:
$$
	\varsigma(y)= \sigma(x)\Partransport\left(x,y\right),
	~~\Partransport\left(x,y\right)\in\iG,
	~~\sigma(x),\varsigma(y)\in\lS;
$$
the connection $\Partransport\left(x,y\right)$ has the obvious property 
	 $\Partransport\left(x,y\right)=\Partransport\left(y,x\right)^{-1};$
	\item 
a connection $\widetilde{\Partransport}\left(x,y\right)$ is called \emph{trivial} 
	if it can be expressed 
	in terms of a function on \emph{vertices} of the graph: 
	$\widetilde{\Partransport}\left(x,y\right)=
	p\left(x\right)p\left(y\right)^{-1},
	~~p\left(x\right),p\left(y\right)\in\iG;$
	\item invariance  with respect to the gauge symmetries 
	depending on time and  space   
	leads to the transformation rule for connection
\begin{equation}
		\Partransport(x,y) \rightarrow \gamma(x)^{-1}\Partransport(x,y)\gamma(y),~~
		\gamma(x), \gamma(y)\in\iG;\label{ginv}
\end{equation}
  \item the \emph{curvature} of connection $\Partransport(x,y)$ 
  is defined as the conjugacy class%
\footnote{The conjugacy equivalence 
 means that 
$\Partransport'(x_1,\ldots,x_k)\sim\gamma^{-1}\Partransport(x_1,\ldots,x_k)\gamma$
for any  $\gamma\in\iG$.}  
of the \emph{ho\-lo\-nomy} along a cycle of a graph: 
    $$
\Partransport(x_1,x_2,\ldots,x_k)=\Partransport(x_1,x_2)\Partransport(x_2,x_3)
\cdots \Partransport(x_k,x_1);
    $$
   the curvature of trivial connection is obviously trivial:
   $\widetilde{\Partransport}(x_1,\ldots,x_k)=\id;$
   \item the gauge principle does not tell us anything about the 
   evolution of the connection
   itself, so gauge invariant relation describing dynamics of connection 
   (\emph{gauge field}) should be added.
\end{itemize}
Let us give two illustrations of how these concepts work in continuous case.
\subsubsection*{Electrodynamics. Abelian prototype of all gauge theories.}
Here the set $\SpaceTime$ is 4-dimensional Minkowski space with points 
$x=\left(x^\mu\right)$ 
and the set of states is Hilbert space of complex scalar 
(Schr\"{o}dinger equation)
or spinor (Dirac equation) fields $\psi(x).$ The symmetry group of the 
Lagrangians and physical 
observables is the unitary group $\iG=\U(1).$ The elements of $\iG^\X$ can be represented
as $e^{-i\alpha(x)}.$ 
\par
Let us  consider the 
parallel transport for two closely situated space-time points: 
$$\Partransport(x, x+\Delta x) = e^{-i\rho(x, x+\Delta x)}.$$
Specializing transformation rule \eqref{ginv} to this particular case
$$
\Partransport'(x, x+\Delta x) 
= e^{i\alpha(x)}\Partransport(x, x+\Delta x)e^{-i\alpha(x+\Delta x)},
$$ 
substituting approximations
$$\Partransport(x, x+\Delta x) = e^{-i\rho(x, x+\Delta x)}\approx\id-i{A}(x)\Delta x,$$
$$\Partransport'(x, x+\Delta x) = e^{-i\rho(x, x+\Delta x)}\approx\id-i{A'}(x)\Delta x,$$
$$e^{-i\alpha(x+\Delta x)}\approx e^{-i\alpha(x)}\left(\id-i\nabla\alpha(x)\Delta x\right),$$
and taking into account commutativity of $\iG=\U(1)$ we obtain
\begin{equation}
{A'}(x)={A}(x)+\nabla\alpha(x)\enspace~~
\mbox{or, in components,}\enspace~~
{A'_\mu}(x)={A_\mu}(x)
+\frac{\textstyle{\partial\alpha(x)}}{\textstyle{\partial x^\mu}}.\label{grad}	
\end{equation}
The 1-form $A$ taking values in the Lie algebra of $\U(1)$ and its differential
$F=\left(F_{\mu\nu}\right)=\mathrm{d}A$ are identified
with the electromagnetic \emph{vector potential} and  the \emph{field strength},
respectively. To provide the gauge invariance of the equations for field $\psi(x)$ 
we should replace
partial  by covariant derivatives
$$
\partial_\mu \rightarrow D_\mu=\partial_\mu-iA_\mu(x)
$$
in those equations.
\par 
Finally, evolution equations  for the gauge field $A(x)$ should be added. 
In the case of 
electromagnetics these are Maxwell's equations:
\begin{eqnarray}
\mathrm{d}F&=&0\hspace*{20pt}\text{\emph{first pair}}\label{m1st}\\	
\mathrm{d}\star F&=&0\hspace*{20pt}\text{\emph{second pair}}\label{m2nd}.	
\end{eqnarray}
Here $\star$ is the \emph{Hodge conjugation} (\emph{Hodge star operator}).
Note that equation \eqref{m2nd} corresponds to \emph{vacuum Maxwell's equations}. 
In the presence of the  
\emph{current} $J~$ the \emph{second pair} takes the form $~~\star\mathrm{d}\star F=J.$ 
Note also that the \emph{first pair}
is essentially \emph{a priori} statement, it reflects simply the fact that $F$, 
by definition, is the 
differential of an exterior form.
\subsubsection*{Non-Abelian Gauge Theories in Continuous Space-time.} 
Only minor modifications are needed for the case of non-Abelian Lie group $\iG$.
Again expansion of the $\iG$-valued parallel transport for two close space-time points $x$ 
and $x+\Delta x$ with taking into account
that $\Partransport(x,x)=1$ leads to introduction of a Lie algebra valued 1-form $A=\left(A_\mu\right):$
$$
\Partransport(x,x+\Delta x)\approx\id+A_\mu(x)\Delta x^\mu.
$$
Infinitesimal manipulations with formula \eqref{ginv}
$$
\gamma(x)^{-1}\Partransport(x,x+\Delta x)\gamma(x+\Delta x)~ \longrightarrow~
\gamma(x)^{-1}\left(\id+A_\mu(x)\Delta x^\mu\right)
\left(\gamma(x)+\frac{\textstyle{\partial \gamma(x)}}{\textstyle{\partial x^\mu}}\Delta x^\mu\right)
$$
lead to the following transformation rule
\begin{equation}
{A'_\mu}(x)=\gamma(x)^{-1}{A_\mu}(x)\gamma(x)
+\gamma(x)^{-1}\frac{\textstyle{\partial \gamma(x)}}{\textstyle{\partial x^\mu}}.\label{nagrad}	
\end{equation}
The curvature 2-form
$$
F = dA+\left[A\wedge A\right]
$$
is interpreted as the \emph{physical strength field}. 
In particular, the \emph{trivial} connection
 $$\widetilde{A}_\mu(x)=\gamma_0(x)^{-1}\frac{\textstyle{\partial \gamma_0(x)}}{\textstyle{\partial x^\mu}}$$ 
is \emph{flat}, i.e., its curvature $F=0.$ 
\par
There are different approaches to construct dynamical equations for gauge fields
\cite{Oeckl}. The most important example is \emph{Yang-Mills theory} based on the Lagrangian
$$
L_{YM}=\mathrm{Tr}\left[F\wedge\star F\right].
$$
The Yang-Mills equations of motion read 
\begin{eqnarray}
\mathrm{d}F+\left[A\wedge F\right]&=&0,\label{Bianci}\\
\mathrm{d}\star F+\left[A\wedge\star F\right]&=&0.
\end{eqnarray}
Here again equation \eqref{Bianci} is  \emph{a priori} statement --- 
the \emph{Bianci identity}. 
Note that Maxwell's equations are a special case of Yang-Mills equations. 
\par
It is instructive to see what the Yang-Mills Lagrangian looks like in the discrete approximation.
Replacing the Minkowski space $\SpaceTime$ by a hypercubic lattice one can see that the discrete
version of $L_{YM}$ is proportional to $\sum_f\sigma\left(\gamma_f\right)$,
where the summation is performed over all faces of a hypercubic constituent of the lattice;
$$
\sigma = 2\dim\Rep{\iG}-\Char{\Rep{\iG}}-\Char{\RepH{\iG}};
$$ 
where $\Rep{\iG}$ and $\RepH{\iG}$ are fundamental representation of $\iG$ and its dual,
respectively;
$\Charbare$ is the character; $\gamma_f$ is the gauge group holonomy around the face $f$.
\par
The Yang-Mills theory 
uses  Hodge operation converting $k$-forms to $(n-k)$-forms
in $n$-dimensional space \emph{with metric} $g_{\mu\nu}$.
In topological applications so-called \emph{BF theory} plays
an important role since it does not require a metric. 
In this theory, an additional dynamical field $B$
is introduced. The Lie algebra valued $(n-2)$-form $B$ and the $2$-form
$F$ are combined into the  Lagrangian $L_{BF}=\mathrm{Tr}\left[B\wedge F\right].$
\subsection{Quantum Behavior and Gauge Connection} 
\begin{figure}[!h]
\centering
\includegraphics[width=0.85\textwidth]{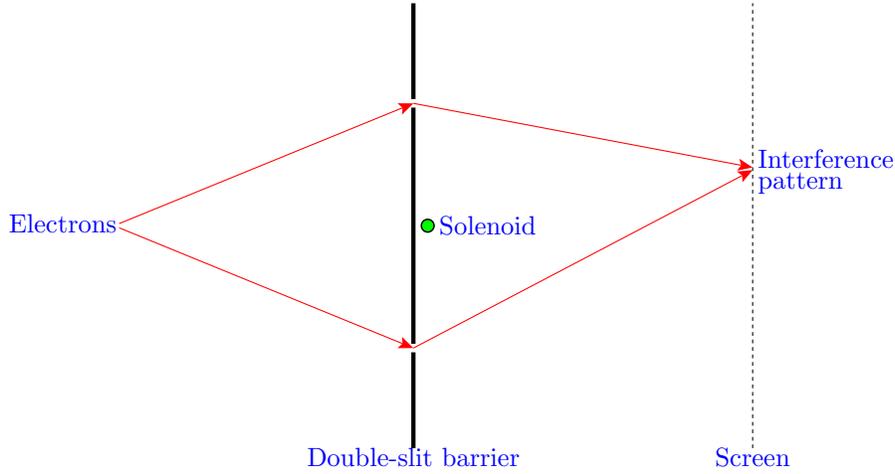}
\caption{Aharonov--Bohm effect. Magnetic flux is confined within the perfectly shielded solenoid;
interference pattern is shifted in spite 
of absence of electromagnetic forces acting on the particles.}
	\label{Aharonov-Bohm}
\end{figure}

The Aharonov--Bohm effect (Fig. \ref{Aharonov-Bohm}) is one of the most remarkable 
illustrations of interplay between quantum behavior and gauge connection. 
Charged particles moving through 
the region containing perfectly shielded thin solenoid 
produce different interference patterns on a screen depending on whether the solenoid 
is turned on or off. 
There is no electromagnetic force acting on the particles, but working solenoid produces 
$\U(1)$-connection adding or subtracting phases of the particles and thus changing the  
interference pattern. 
\par
In the discrete time Feynman's path amplitude  \cite{Feynman} is decomposed into 
the product of elements of the fundamental representation $\Rep{\iG}=\U(1)$ 
of the circle, i.e.,  of
the Lie group $\iG=\mathrm{S}^1=\R/\Z$:
\begin{equation}
	 A_{\U(1)}=\exp\left(iS\right)=\exp\left(i\int Ldt \right)\longrightarrow
	  \e^{\textstyle{iL_{0,1}}}\ldots\e^{\textstyle{iL_{t-1,t}}}
	  \ldots\e^{\textstyle{iL_{T-1,T}}}.\label{FPA}
\end{equation}
By the notation $L_{t-1,t}$  we emphasize that the Lagrangian is in fact a function
defined on pairs of points (graph edges) --- this is compatible with physics 
where the typical Lagrangians are depend on the \emph{first order} derivatives. 
Thus we can interpret the expression
 $\Partransport(t-1,t)=e^{\textstyle{iL_{t-1,t}}}\in\Rep{\iG}=\U(1)$ 
 as $\U(1)$-parallel transport.
\par
A natural generalization of this is
to suppose that: 
\begin{itemize}
\item group $\iG$ may differ from  $\mathrm{S}^1$, 
\item dimension of unitary representation $\Rep{\iG}$ may differ from 1.
\end{itemize}
So let us replace expression \eqref{FPA} for Feynman's path amplitude  by the following 
parallel transport along the path
\begin{equation}
	 A_{\Rep{\iG}}=\Rep{\alpha_{T,T-1}}\ldots\Rep{\alpha_{t,t-1}}
	  \ldots\Rep{\alpha_{1,0}}.\label{Gammaampl}
\end{equation}
Here $\alpha_{t,t-1}$ are elements of some group
$\iG$ --- we shall call it \emph{quantizing group} --- and $\Repbare$ is 
an unitary representation of $\iG$.
Note that in \eqref{FPA} the order of factors is not important due to commutativity of $\U(1)$.
But in \eqref{Gammaampl}  we must use the reverse%
\footnote{This awkwardness stems from the tradition to write operator 
actions on the left (cf. footnote~\ref{ontheright} on page \pageref{ontherightpage}).}
order for consistency with the temporal ordering of non-commutative operators.
For discrete and especially finite systems it is natural 
to take a finite group as the quantizing group, in this case all  
manipulations --- in contrast to the standard quantization --- remain within the framework
of constructive discrete mathematics requiring no more than the ring of
\emph{algebraic integers}  (and sometimes the quotient field of this ring). 
On the other hand, the standard quantization can be approximated by taking 
1-dimensional representations of large enough finite groups.

\subsubsection{Illustrative Example Inspired by Free Particle.}
In quantum mechanics --- as is clear from the \emph{never vanishing} expression 
$\exp\left(\frac{i}{\hbar}S\right)$ for the path 
amplitude --- transitions from one to any other state are possible in principle.
But we shall consider computationally more tractable models with restricted 
sets of possible transitions.
\par
Let us consider quantization of a free particle moving in one dimension. 
Such a particle is described
by the Lagrangian $L = \frac{\textstyle{m\dot{x}^2}}{\textstyle{2}}.$ 
Assuming that there are only transitions to the closest points in the discretized space
we come to the following rule for the one-time-step transition amplitudes
\begin{center}
\includegraphics[width=0.20\textwidth]{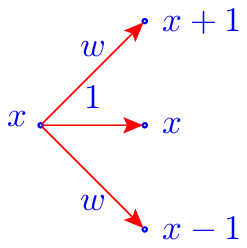}\hspace*{20pt}
\raisebox{34pt}{\begin{tabular}{l}
$e^{\textstyle{\frac{i}{\hbar}\frac{m\left\{(x+1)-x\right\}^2}{2}}}
=e^{\textstyle{i\frac{m}{2\hbar}}}$\\[10pt]
$e^{\textstyle{\frac{i}{\hbar}\frac{m\left(x-x\right)^2}{2}}}\hspace*{17pt}=~\id$\\[10pt]
$e^{\textstyle{\frac{i}{\hbar}\frac{m\left\{(x-1)-x\right\}^2}{2}}}=e^{\textstyle{i\frac{m}{2\hbar}}}.$
\end{tabular}}
\end{center}
That is, we have evolution rule as an $\U(1)$-valued function $R$ defined on pairs of points (graph edges).
Symbolically: 
\begin{eqnarray}
	R\left(x\rightarrow{}x\right)&=&1~~\in~~\U(1),\nonumber\\
	R\left(x\rightarrow{}x-1\right)=R\left(x\rightarrow{}x+1\right)
	&=&w=e^{\textstyle{i\frac{m}{2\hbar}}}~~\in~~\U(1).
	\label{rfree}
\end{eqnarray}
Now let us assume that $w$ in \eqref{rfree} is an element of some representation 
of a finite group: $w=\Rep{\alpha},~\alpha\in\iG=\set{\gamma_1=\id,\ldots,\gamma_{M}}$.
Rearranging \emph{multinomial coefficients} --- \emph{trinomial} in this concrete case 
--- it is not difficult to write the sum amplitude over all paths from the space-time point 
$\left(0,0\right)$ to the point $\left(x,t\right)$
\begin{equation}
A_x^t\left(w\right)=\sum\limits_{\tau=0}^t\frac{\tau!}{\left(\frac{\tau-x}{2}\right)!\left(\frac{\tau+x}{2}\right)!}
	\times
	\frac{t!}{\tau!\left(t-\tau\right)!}~w^{\tau}.
\end{equation}
Note that $x$ must lie in the limits determined by $t$: $x\in\left[-t,t\right].$
\par
One of the most expressive peculiarities of quantum-mechanical behavior is the 
\emph{destructive interference} --- cancellation of non-zero amplitudes attached
to different paths converging to the same point. 
By construction, the sum of amplitudes in our model is a function $A(w)$ depending 
on distribution of sources of the particles, their initial phases,  gauge fields 
acting along the paths, restrictions --- like, e.g., ``slits'' --- imposed on possible 
paths, etc.
In the case of 1-dimensional representation the function $A(w)$ is a polynomial with algebraic 
integer coefficients and $w$ is a root of unity. Thus the condition for destructive interference
can be expressed by the system of polynomial equations: $A(w)=0$ and $w^M=1$.
For concreteness let us consider the cyclic group 
$\iG=\Cgr{M}=\set{\gamma_1,\cdots,\gamma_k,\cdots,\gamma_{M}}$. 
Any of its $M$ irreducible representations takes the form $\Rep{\gamma_k}=w^{k-1}$, 
where $w$ is one of the $M$th roots of unity.
For simplicity let $w$ be the \emph{primitive root}: $w=\e^{2\pi{}i/M}.$
\begin{figure}[!h]
\centering
\includegraphics[width=0.95\textwidth]{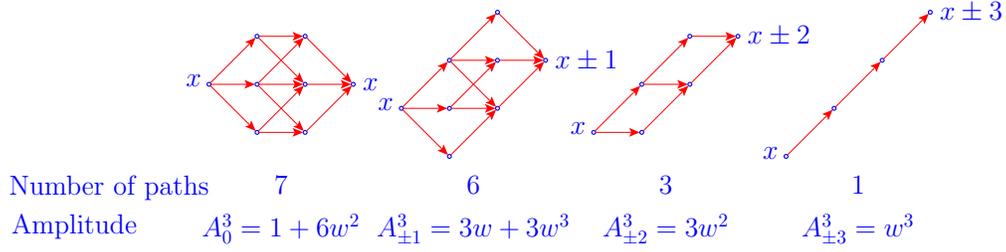}
\caption{Amplitudes for all possible paths in three time steps.}
	\label{Ampl3T}
\end{figure}
\par
Fig. \ref{Ampl3T} shows all possible transitions (with their amplitudes) from the point $x$ 
in three time steps.
We see that the polynomial $A^3_{\pm1}=3w+3w^3=3w\left(w^2+1\right)$ contains the
\emph{cyclotomic polynomial} $\Phi_4(w)=w^2+1$ as a factor. 
The smallest group associated to $\Phi_4(w)$ --- and hence providing the destructive 
interference --- is $\Cgr{4}$, 
which we shall consider as quantizing group for the model.
\begin{figure}[!h]
\centering
\includegraphics[width=0.95\textwidth]{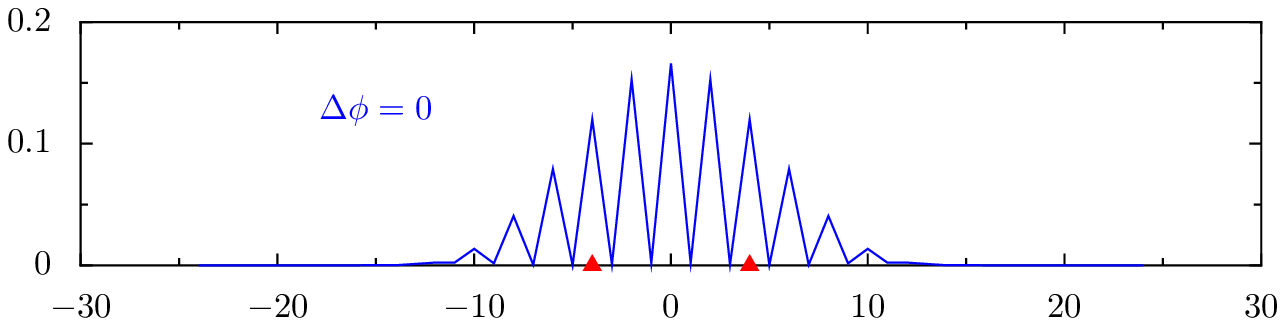}\\
\includegraphics[width=0.95\textwidth]{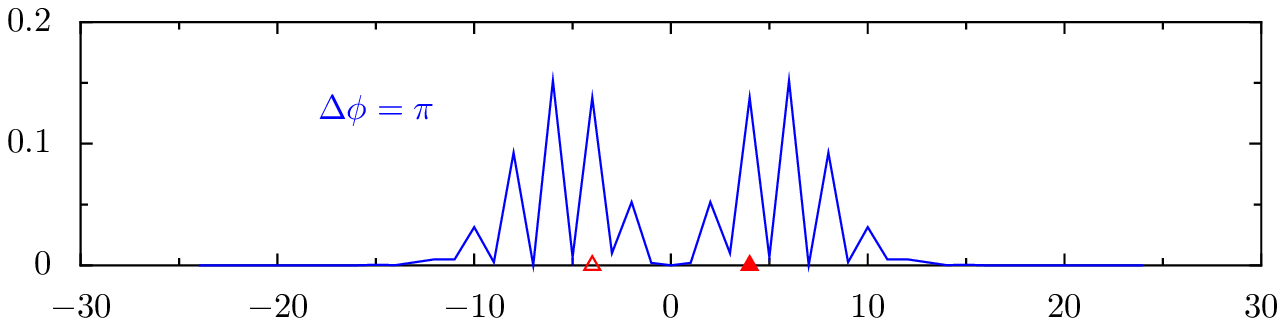}
\caption{Group $\Cgr{4}$. Interference from two sources at points -4 and 4. 
Number of time steps $T=20$. Phase 
differences $\Delta\phi=\phi_4-\phi_{-4}$ between
sources are $0$ and $\pi$.}
	\label{Interf2}
\end{figure}
\par
Fig. \ref{Interf2} shows interference patterns --- normalized squared amplitudes 
(``probabilities'') ---  from two sources placed in the positions $x=-4$ and $x=4$ 
for 20 time steps. The upper and lower graph show interference pattern when sources
are in the same ($\Delta\phi=0$) and in the opposite 
($\Delta\phi=\pi$) phases, respectively.
\subsubsection{Local Quantum Models on Regular Graphs.}
The above model --- with quantum transitions allowed only within the neighborhood 
of a vertex
of a 1-dimensional lattice --- can easily be generalized to arbitrary regular graph.
Our definition 
of \emph{local quantum model on $k$-valent graph} uncludes the following:
\begin{enumerate}
\item
 \emph{Space} $X=\set{x_1,\cdots,x_{N_\X}}$ is a $k-$valent graph.
\item
 \emph{Set of local transitions} $E_i=\set{e_{0,i}, e_{1,i},\cdots,e_{k,i}}$ 
 is the set of $k$ adjacent to the vertex $x_i$ edges 	$e_{m,i}=\left(x_i\rightarrow{}x_{m,i}\right)$
 completed by the edge	$e_{0,i}=\left(x_i\rightarrow{}x_i\right)$.
\item
 We assume that the \emph{space symmetry} group $\sG=\mathrm{Aut}\left(\X\right)$ acts transitively
 on the set $\set{E_1,\cdots,E_{N_\X}}$.
\item
 $\sGloc=\mathrm{Stab}_{\sG}\left(x_i\right)\leq\sG$ is the \emph{stabilizer} of $x_i$.
\item
 $\Omega_i=\set{\omega_{0,i},\omega_{1,i},\cdots,\omega_{h,i}}$	
	is the \emph{set of orbits} of $\sGloc$ on $E_i$.
\item
 \emph{Quantizing group} $\iG$ is a finite group: $\iG=\set{\gamma_1,\cdots,\gamma_{M}}$.
\item
 \emph{Evolution rule} $R$ is a function on $E_i$ with values in some 
 representation $\Rep{\iG}$. The rule $R$ prescribes $\Rep{\iG}$-weights 
	to the one-time-step transitions from $x_i$ to elements of the neighborhood of $x_i$.
	From the symmetry considerations $R$ must be a function on orbits from
	$\Omega_i$, i.e., $R\rbra{e_{m,i}g}=R\rbra{e_{m,i}}$ for $g\in\sGloc$.
\end{enumerate}
To illustrate these constructions, let us consider the local quantum model on the graph 
of \emph{buckyball} (see detailed consideration of this graph at page \pageref{buckyball}). 
Here the space $X=\set{x_1,\cdots,x_{60}}$ has the  symmetry group  
$\sG=\mathrm{Aut}\left(\X\right)=\Cgr{2}\times\Alt{5}$.
The set of local transitions takes the form $E_i=\set{e_{0,i},~ e_{1,i},~ e_{2,i},~ e_{3,i}}$, 
where~ 
\begin{eqnarray*}
	 e_{0,i}&=&\rbra{x_i\rightarrow{}x_i},\\
	 e_{1,i}&=&\rbra{x_i\rightarrow{}x_{1,i}},\\
	 e_{2,i}&=&\rbra{x_i\rightarrow{}x_{2,i}},\\
	 e_{3,i}&=&\rbra{x_i\rightarrow{}x_{3,i}}.
\end{eqnarray*}
The stabilizer of $x_i$ is $\sGloc=\mathrm{Stab}_{\sG}\left(x_i\right)=\Cgr{2}$. 
The set of orbits of $\sGloc$ on $E_i$ contains 3 orbits:
 $$\Omega_i=\set{\omega_{0,i}=\set{e_{0,i}}, \omega_{1,i}
=\set{e_{1,i}, e_{2,i}}, \omega_{2,i}=\set{e_{3,i}}},$$
 i.e.,
the stabilizer does not move the edges $\rbra{x_i\rightarrow{}x_i}$
 and
 $\rbra{x_i\rightarrow{}x_{3,i}}$ and\\ swaps $\rbra{x_i\rightarrow{}x_{1,i}}$ and
 $\rbra{x_i\rightarrow{}x_{2,i}}.$ 
 \par
The evolution rule takes the form:
\begin{eqnarray*}
R\rbra{x_i\rightarrow{}x_i}&=&\Rep{\alpha_0},\\
R\rbra{x_i\rightarrow{}x_{1,i}}=
R\rbra{x_i\rightarrow{}x_{2,i}}&=&\Rep{\alpha_1},\\
R\rbra{x_i\rightarrow{}x_{3,i}}&=&\Rep{\alpha_2},	
\end{eqnarray*}
where $\alpha_0,\alpha_1,\alpha_2\in\iG$. If we take a 1-dimensional 
representation and move 
$\alpha_0$ --- using gauge invariance --- to the identity element of $\iG$, 
we see that the rule $R$ depends
on  two elements $v=\Rep{\alpha_1}$ and  $w=\Rep{\alpha_2}$. 
Thus the amplitudes in the quantum model on the buckyball take the form $A(v, w)$ 
depending on two roots of unity. To search nontrivial quantizing groups one should check 
(by, e.g., Gr\"{o}bner basis computation) compatibility of the system of 
polynomial equations
$A(v, w)=\Phi_i(v)=\Phi_j(w)=0$, where $\Phi_i(v)$ and $\Phi_j(w)$ are cyclotomic polynomials.
\subsection{General Discussion of Quantization in Finite Systems}
As is well known, Feynman's approach is equivalent to the traditional
matrix formulation of quantum mechanics, where the time evolution 
$\left|\psi_0\right\rangle\rightarrow\left|\psi_T\right\rangle$ of a system from 
the initial state vector to the final is described by the evolution
matrix $U$: $\left|\psi_T\right\rangle=U\left|\psi_0\right\rangle$.
The evolution matrix can be represented as the product of matrices corresponding to
the single time steps: 
$U=U_{T\leftarrow{}T-1}\cdots{}U_{t\leftarrow{}t-1}\cdots{}
U_{1\leftarrow{}0}$.
In fact, Feynman's quantization --- i.e., the rules ``multiply subsequent events'' and 
``sum up alternative histories'' --- is simply a rephrasing of matrix multiplication.
This is clear from the below illustration presenting  two-time-step evolution of a 
two-state system (single \emph{qubit}) in both Feynman's and matrix forms --- 
the general case of many time steps and many states can easily be obtained (by 
induction for example).
\begin{center}
\begin{tabular}[t]{ccc}
\includegraphics[height=0.24\textwidth]{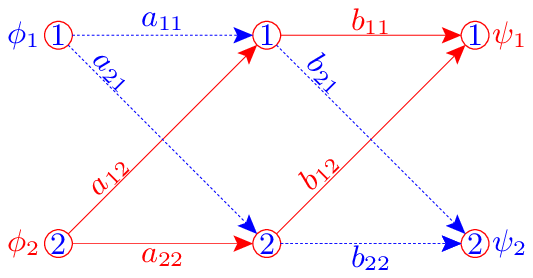}
&\raisebox{0.1\textwidth}{$\sim$}
&
\includegraphics[height=0.24\textwidth]{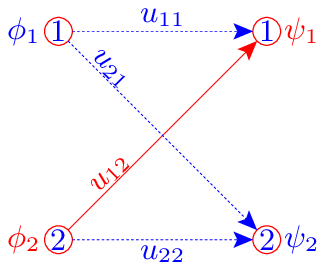}
\\
$\Updownarrow$&&$\Updownarrow$
\\
$
BA=
\begin{bmatrix}
{\color{blue}b_{11}a_{11}+b_{12}a_{21}}
&
{\color{red}b_{11}a_{12}+b_{12}a_{22}}
\\[5pt]
{\color{blue}b_{21}a_{11}+b_{22}a_{21}}
&
{\color{blue}b_{21}a_{12}+b_{22}a_{22}}
\end{bmatrix}
$~~~~~~~~~
& $\sim$~~ &
$
U=
\begin{bmatrix}
{\color{blue}u_{11}}&{\color{red}u_{12}}
\\[5pt]
{\color{blue}u_{21}}&{\color{blue}u_{22}}
\end{bmatrix}
$~~~~~~~~
\end{tabular}
\end{center}
We see that in accordance with Feynman's rules the transition from, e.g., $\phi_2$ to $\psi_1$
is determined by the expression $b_{11}a_{12}+b_{12}a_{22}$. 
But this is just the element $u_{12}$
of the matrix product $U=BA$ performing  evolution $\left|\psi\right\rangle=U\left|\phi\right\rangle$, 
where   
$\left|\phi\right\rangle=
\begin{bmatrix}
\phi_{1}\\
\phi_{2}
\end{bmatrix}
$ 
and 
$\left|\psi\right\rangle=
\begin{bmatrix}
\psi_1\\
\psi_2
\end{bmatrix}
$.
\par
Of course, such reduction of sums over histories to matrices is applicable to the case 
of transitions along paths being gauge connections as in \eqref{Gammaampl}. In this
case matrix elements of an $\XN\times\XN$ evolution matrix $U$ are themselves matrices
from the representation $\Rep{\iG}$. We can ignore this particular block structure of 
the matrix and consider $U$ as an $N\times{}N$ matrix over the field $\C$, 
where $N=\XN\times\dim\Rep{\iG}$.
\par
In quantum mechanics, the evolution matrices $U$ are unitary operators acting in Hilbert 
spaces of state vectors (called also ``wave functions'', ``amplitudes'' etc.). 
Quantum mechanical particles are associated with unitary representations of some groups.
According to their dimensions, these representations are called ``singlets'', ``doublets'', etc.
Multidimensional representations describe the \emph{spin}. A quantum mechanical experiment
is reduced to comparison of the system state vector $\left|\psi\right\rangle$ with 
some sample state vector $\left|\phi\right\rangle$. 
According to the Born rule, the probability to observe coincidence of the states 
is equal to $\left|\left\langle\phi|\psi\right\rangle\right|^2$, where 
$\left\langle\cdot|\cdot\right\rangle$ is the inner product in the Hilbert space.
To see what these constructions may look like in the constructive finite background, 
let us assume that evolution operators are elements of a representation of a finite group.
\subsubsection{Permutations and Linear Representations}
Having a finite group $G=\set{e_1=\id,\ldots,e_m}$, we can easily describe all its transitive
actions on finite sets \cite{Hall}. 
Any such set $\Omega=\set{\omega_1,\ldots,\omega_n}$ is in one-to-one correspondence
with the \emph{right} (or \emph{left}) \emph{cosets} of some subgroup $H\leq{}G$, i.e., 
$\Omega\cong{}H\setminus{}G~\left(\text{or~~}G/H\right)$ is the \emph{homogeneous space}
(or $G$-\emph{space}). 
Action of $G$ on $\Omega$ is \emph{faithful} if the subgroup $H$ does not contain
\emph{normal sugroups} of $G$. We can write actions in terms of permutations
$$
	\pi(g)=\dbinom{\omega_i}{\omega_ig}\sim\dbinom{Ha}{Hag},
	\hspace*{20pt}g,a\in{}G,~~~i=1,\ldots,n.
$$
Maximum transitive set $\Omega$ is the group itself, i.e., in the above construction
 $H=\set{\id}$. The action on $\Omega=G$ is called \emph{regular} and can be represented 
 by permutations
\begin{equation}
\Pi(g)=\dbinom{e_i}{e_ig},~~~i=1,\ldots,m.
\label{regact}
\end{equation}
To introduce ``numerical'' (``statistical'') description, let us assume that $\omega_i$'s are basis
elements of a linear vector space $\Hspace$ over a field $F$
\begin{equation}
	\Hspace=\mathrm{Span}\vect{\omega_1,\cdots,\omega_n},
	\label{hspacedef}
\end{equation}
i.e., we prescribe 
$F$-valued ``weights'' to the elements $\omega_i\in\Omega$.
Then we can write permutations in the matrix form: 
\begin{equation}
	\pi(g)\rightarrow\rho(g)=
	\begin{bmatrix}
	\rho(g)_{ij}
	\end{bmatrix}
	,\text{~~ where~~} \rho(g)_{ij}=\delta_{\omega_ig,\omega_j};~~ i,j=1,\ldots,n;
\label{permrep}
\end{equation}
$$
\delta_{\alpha,\beta}\equiv
\begin{cases}
1, & \text{if~~} \alpha=\beta,\\
0, & \text{if~~} \alpha\neq\beta
\end{cases}
\text{~~for~~} \alpha,\beta\in\Omega.
$$ 
The so defined function $\rho$ is called a \emph{permutation representation}.
The matrix form of \eqref{regact} 
\begin{equation}
	\Pi(g)\rightarrow{}\regrep(g)=
	\begin{bmatrix}
	\regrep(g)_{ij}
	\end{bmatrix}
	,~~ \regrep(g)_{ij}=\delta_{e_ig,e_j},~~ i,j=1,\ldots,m
	\label{regrep}
\end{equation}
is called 
the \emph{regular representation}. 
It is assumed that $F$ is an algebraically closed field --- usually 
the field of complex numbers $\C$. 
But in the case of finite groups the quotient field of the ring $\A$ of 
\emph{algebraic integers}%
\footnote{The ring of algebraic integers consists of the roots
of \emph{monic} polynomials with integer coefficients.
A polynomial is called \emph{monic} if its leading coefficient is unit.} 
\cite{Kirillov} 
is sufficient for all reasonable purposes ---
$\A$ is a \emph{constructive} subset of $\C$.
\par 
Let us recall some relevant background information about linear representations 
of finite groups \cite{Serre}.
\begin{enumerate}
\item 
\emph{Any linear} representation of a finite group $G$ is \emph{unitary}
since there is always an unique invariant inner product $\left\langle\cdot|\cdot\right\rangle$
making any space of representation 
 $\Hspace$  into  a Hilbert space.
\item
\emph{All possible} irreducible unitary representations of the group $G$ are contained in the
regular representation \eqref{regrep}. More specifically, all matrices \eqref{regrep} can
simultaneously be reduced
by some unitary transformation $\transmatr$ to the form
\begin{equation}
	\transmatr^{-1}\regrep(g)\transmatr=
		\begin{bmatrix}
		\irrep_1(g) &&&
\\[5pt]
&
\hspace*{-27pt}d_2\left\{
\begin{matrix}
\irrep_2(g)&&\\
&\hspace*{-10pt}\ddots&\\
&&\hspace*{-7pt}\irrep_2(g)
\end{matrix}
\right. 
&&
\\
&&\hspace*{-10pt}\ddots&\\
&&& 
\hspace*{-25pt}d_r\left\{
\begin{matrix}
\irrep_r(g)&&\\
&\hspace*{-10pt}\ddots&\\
&&\hspace*{-7pt}\irrep_r(g)
\end{matrix}
\right. 
		\end{bmatrix}.
		\label{regrepdecomposed}
\end{equation}
Here $r$ is the number of different irreducible representations $\irrep_j$ of the group $G$. This
number coincides with the number of \emph{conjugacy classes}%
\footnote{The $j$th conjugacy class $C_j\subseteq{}G$ consists of all group elements of 
the form $g^{-1}c_jg$, where $c_j\in{}C_j$ is some (arbitrary) representative of the class,
$g\in{}G$, $j=1,\ldots,r$.}
in $G$. The number $d_j$ is simultaneously the dimension of $\irrep_j$ and its multiplicity in 
the regular representation, so it is obvious that $d_1^2+d_2^2+\cdots+d_r^2=|G|=m$. It can be 
proved also that any $d_j$ divides the number of elements of $G$: $d_j\mid{}m$.
	\item 
Any irreducible representation $\irrep_j$ is determined 
uniquely (up to isomorphism) by its \emph{character} $\chi_j$. 
The character is a function on $G$ defined as
$\chi_j(a)=\Tr\irrep_j(a),~a\in{}G.$ The character is a \emph{central} or \emph{class} function, i.e., it is constant
on the conjugacy classes: $\chi_j(a)=\chi_j(g^{-1}ag),~a,g\in{}G$.
Any class function  $\varphi(a)$ on $G$ is a linear combination of the characters
$\chi_{1},\ldots,\chi_{r}$.
\item All values of $\chi_j$ and eigenvalues of $\irrep_j$ are elements of 
the ring  $\A$ of {algebraic integers},{~} moreover{~} the eigenvalues are \emph{roots of unity.}
\item 
A convenient form of describing all irreducible representation of a finite group $G$ is 
 the \emph{character table}. The columns of this table correspond to the conjugacy classes
of $G$ while its rows correspond to the characters $\chi_j$ of the inequivalent 
irreducible representations of $G$.
\begin{center}
\begin{tabular}{c|cccc}
&$\id$&$c_2$&$\cdots$&$c_r$\\\hline
$\chi_1$&1&1&$\cdots$&1\\
$\chi_2$&$\chi_2(c_1)$&$\chi_2(c_2)$&$\cdots$&$\chi_2(c_r)$\\
$\vdots$&$\vdots$&$\vdots$&&$\vdots$\\
$\chi_r$&$\chi_r(c_1)$&$\chi_r(c_2)$&$\cdots$&$\chi_r(c_r)$
\end{tabular}
\end{center}
The $j$th column is indicated by a representative $c_j\in{}C_j$ of the $j$th conjugacy class $C_j$.
Conventionally we take $c_1=\id$ and $\chi_1$ to be the \emph{trivial character} 
corresponding to the trivial 1-dimensional representation.
\end{enumerate}
\subsubsection{Interpretation of Quantum Description in Finite Background}
Let us discuss sketchy (more detailed presentation see in \cite{KornyakFQM10}) 
constructive approach to the interpretation of  quantum description.
\par
Summarizing the above, 
we see that dynamics of finite quantum model of any type is
reduced ultimately to  a single finite-dimensional unitary $k\times{}k$ matrix $U$ describing 
transitions between initial and final vectors in some $k$-dimensional Hilbert space
$\Hspace_k$. In the finite background the matrix $U$ is an element of unitary representation
$\Delta$ of a finite group $G$, i.e., the number of all possible evolutions is equal to $m=|G|$.
We shall assume, as is accepted in quantum mechanics, that $\Delta$ is direct sum of irreducible
representations $\Delta_j$ from \eqref{regrepdecomposed}.
The decomposition of the Hilbert space into irreducible components is an important part
of the mathematical formulation of quantum mechanics. Such dependence on the choice of the 
basis in the Hilbert space may seem unusual for a physical theory. 
But, in fact, a basis in which the Hilbet space is reduced --- we shall call such a basis 
\emph{quantum basis} --- simply reflects the structure of underlying symmetry group. 
\par
We can construct a $G$-space $\Omega=\set{\omega_1,\ldots,\omega_n}$%
\footnote{In the case that $\Delta$ is reducible representation, the set $\Omega$ may be
intransitive union of transitive $G$-spaces.}
in such a way that its permutation representation \eqref{permrep} contains $\Delta$ 
as subrepresentation (obviously $n\geq{}k$). 
That is, the space $\Hspace_k$ is subspace of the Hilbert space $\Hspace_n$ of the 
permutation representation. 
We shall call the basis $\set{\omega_1,\ldots,\omega_n}$ in the space $\Hspace_n$ 
the \emph{permutation basis}.
Transitions from the permutation to quantum basis for matrices $\widetilde{U}$ and 
vectors $\left|\widetilde{\psi}\right\rangle\in\Hspace_n$ are given by the formulas
\begin{eqnarray}
\widetilde{U}_q&=&S^{-1}\widetilde{U}_pS,\label{transmatrix}
\\
\left|\widetilde{\psi}_q\right\rangle&=&S^{-1}\left|\widetilde{\psi}_p\right\rangle.
\label{transvector}
\end{eqnarray}
Now we can embed any evolution $U$ with the matrix $\Delta$ in the space $\Hspace_k$ into 
the evolution $\widetilde{U}$ in the space $\Hspace_n$. In the quantum basis the matrix of 
$\widetilde{U}$ takes the form 
\begin{equation}
	\widetilde{U}_q=
	\begin{bmatrix}
	\Delta&0\\
	0&A
	\end{bmatrix},
	\label{UQ}
\end{equation}
where $A$ is an $(n-k)\times(n-k)$ matrix. Due to the form of \eqref{UQ} the 
evolution $U$ described by $\Delta$ is completely independent of the components of vectors of 
$\Hspace_n$ related to $A$. The ``hidden variables'' that can come from the additional
components describe degrees of freedom reflecting indistinguishability of $\omega_i$'s
lying in the same group orbit.
The evolution $\widetilde{U}$ is simply a permutation
of $\omega_i$'s and can not manifest anything quantum.
\par
\paragraph{Illustration. A quantum model with the group $\Perm{3}$.}
The group $G=\Perm{3}$ is the group of all permutations of three objects.
This is the smallest non-commutative group. Its 6 elements form the following 
3 conjugate classes $C_1=\set{\id=()}$, $C_2=\set{a_1=(12),a_2=(23),a_3=(13)}$,
$C_3=\set{b_1=(123),b_2=(132)}$. 
We used here the \emph{cyclic notation} for permutations.
The group has three nonequivalent irreducible representations 
described by the character table
\begin{center}
\begin{tabular}{c|crr}
&$\id$&$a_i$&$b_i$\\\hline
$\chi_1$&1&1&1\\
$\chi_2$&1&-1&1\\
$\chi_3$&2&0&-1
\end{tabular}\enspace.
\end{center}
Let us take for example the 2-dimensional representation 
$\Delta$ with the character  $\chi_3$.
The representation is given by the following  set of $2\times2$ matrices:
\begin{center}
$
\Delta(\id)=
	\begin{bmatrix}
	1&0\\
	0&1
	\end{bmatrix},
$
\end{center}
\begin{center}
$	
\Delta(a_1)=
	\begin{bmatrix}
	0&\e^{-2\pi{}i/3}\\
	\e^{2\pi{}i/3}&0
	\end{bmatrix},~~~~
$
$
\Delta(a_2)=
	\begin{bmatrix}
	0&1\\
	1&0
	\end{bmatrix},~~~~
$
$	
\Delta(a_3)=
	\begin{bmatrix}
	0&\e^{2\pi{}i/3}\\
	\e^{-2\pi{}i/3}&0
	\end{bmatrix},
$
\end{center}
\begin{center}
$	
\Delta(b_1)=
	\begin{bmatrix}
	\e^{2\pi{}i/3}&0\\
	0&\e^{-2\pi{}i/3}
	\end{bmatrix},~~~~
$
$	
\Delta(b_2)=
	\begin{bmatrix}
	\e^{-2\pi{}i/3}&0\\
	0&\e^{2\pi{}i/3}
	\end{bmatrix}.
$
\end{center}
The regular permutation representation of $\Perm{3}$ is 6-dimensional. 
But 3-dimensional faithful~ permutation~ representation induced by the action on the 
homogeneous space 
$\Perm{2}\setminus\Perm{3}\cong\Omega=\set{\omega_1,\omega_2,\omega_3}$ also contains $\Delta$.
Since any permutation representation contains trivial 1-dimension subrepresentation,
the only possible choice of the addition $A$ is the representation 
corresponding to the first row of the above character table.
Thus, for $\widetilde{U}_q$ we have
\begin{center}
$
\widetilde{U}_q(\id)=
	\begin{bmatrix}
	1&0&0\\
	0&1&0\\
	0&0&1
	\end{bmatrix},~
$
\end{center}
\begin{center}
$	
\widetilde{U}_q(a_1)=
	\begin{bmatrix}
	0&\!\e^{-2\pi{}i/3}\!&0\\
	\!\e^{2\pi{}i/3}\!&0&0\\
	0&0&1
	\end{bmatrix},
$
$
\widetilde{U}_q(a_2)=
	\begin{bmatrix}
	0&1&0\\
	1&0&0\\
	0&0&1
	\end{bmatrix},
$
$	
\widetilde{U}_q(a_3)=
	\begin{bmatrix}
	0&\!\e^{2\pi{}i/3}\!&0\\
	\!\e^{-2\pi{}i/3}\!&0&0\\
	0&0&1
	\end{bmatrix},~
$
\end{center}
\begin{center}
$	
\widetilde{U}_q(b_1)=
	\begin{bmatrix}
	\e^{2\pi{}i/3}&0&0\\
	0&\e^{-2\pi{}i/3}&0\\
	0&0&1
	\end{bmatrix},~~~~
$
$	
\widetilde{U}_q(b_2)=
	\begin{bmatrix}
	\e^{-2\pi{}i/3}&0&0\\
	0&\e^{2\pi{}i/3}&0\\
	0&0&1
	\end{bmatrix}.
$
\end{center}
In the permutation basis we have
\begin{center}
$
\widetilde{U}_p(\id)=
	\begin{bmatrix}
	1&0&0\\
	0&1&0\\
	0&0&1
	\end{bmatrix},~~~~
$
$	
\widetilde{U}_p(a_1)=
	\begin{bmatrix}
	0&1&0\\
	1&0&0\\
	0&0&1
	\end{bmatrix},~~~~
$
$
\widetilde{U}_p(a_2)=
	\begin{bmatrix}
	1&0&0\\
	0&0&1\\
	0&1&0
	\end{bmatrix},
$
\end{center}
\begin{center}
$	
\widetilde{U}_p(a_3)=
	\begin{bmatrix}
	0&0&1\\
	0&1&0\\
	1&0&0
	\end{bmatrix},~~~~
$
$	
\widetilde{U}_p(b_1)=
	\begin{bmatrix}
	0&1&0\\
	0&0&1\\
	1&0&0
	\end{bmatrix},~~~~
$
$	
\widetilde{U}_p(b_2)=
	\begin{bmatrix}
	0&0&1\\
	1&0&0\\
	0&1&0
	\end{bmatrix}.
$
\end{center}
The most general unitary matrix of transition from the 
permutation to the quantum basis takes the form
\begin{equation}
	S=\frac{\e^{{\alpha{i}}}}{\sqrt{3}}
		\begin{bmatrix}
		1&1&
		\e^{{\beta{i}}}\\
		\e^{{2\pi{}i/3}}&\e^{{-2\pi{}i/3}}&
		\e^{{\beta{i}}}\\
		\e^{{-2\pi{}i/3}}&\e^{{2\pi{}i/3}}&
		\e^{{\beta{i}}}
		\end{bmatrix}, \text{~~~where~~} \alpha, \beta \text{~~are arbitrary real parameters.}
		\label{Smatrix}
\end{equation}
Any quantum evolution of the form
$\left|\psi\right\rangle=U\left|\phi\right\rangle$, 
where   
$\left|\phi\right\rangle=
\begin{bmatrix}
\phi_{1}\\
\phi_{2}
\end{bmatrix}
$ 
and 
$\left|\psi\right\rangle=
\begin{bmatrix}
\psi_1\\
\psi_2
\end{bmatrix}
$, and $U$ is one of the matrices  $\Delta$;
can be extended to the evolution 
$\left|\widetilde{\psi}_q\right\rangle=\widetilde{U}_q\left|\widetilde{\phi}_q\right\rangle$,
where   
$\left|\widetilde{\phi}_q\right\rangle=
\begin{bmatrix}
\phi_{1}\\
\phi_{2}\\
\phi_{3}
\end{bmatrix}
$ 
and 
$\left|\widetilde{\psi}_q\right\rangle=
\begin{bmatrix}
\psi_1\\
\psi_2\\
\psi_3
\end{bmatrix}
$, $\phi_{3}$ is arbitrary additional component. Then, applying the transformation $S$, 
we come to the classical evolution with the matrix 
$\widetilde{U}_p=S\widetilde{U}_qS^{-1}$ which simply permutes 
the components of the initial vector 
$$
\left|\widetilde{\phi}_p\right\rangle=S\left|\widetilde{\phi}_q\right\rangle=
\frac{\e^{{\alpha{i}}}}{\sqrt{3}}
\begin{bmatrix}
\phi_{1}+\phi_{2}+\e^{{\beta{i}}}\phi_{3}\\
\e^{{2\pi{}i/3}}\phi_{1}+\e^{{-2\pi{}i/3}}\phi_{2}+\e^{{\beta{i}}}\phi_{3}\\
\e^{{-2\pi{}i/3}}\phi_{1}+\e^{{2\pi{}i/3}}\phi_{2}+\e^{{\beta{i}}}\phi_{3}
\end{bmatrix}
$$
without performing any algebraic manipulations with the components. 
\section{Conclusion}
In this chapter we discuss the general concept of discrete dynamical system and 
its specialization involving underlying  space structures. We apply various 
constructive approaches to study discrete and finite dynamical systems.
\par 
We construct a family of groups unifying space and internal symmetries in 
 a natural  way. This construction generalizes the standard {direct} and 
 {wreath} products.
\par
We introduce the concept of a {system of
discrete relations on an abstract simplicial complex.}
This system can be treated as a natural generalization of
cellular automata or as a set-theoretical analog of systems
of polynomial equations.
\par
We developed and implemented algorithms for {analyzing
compatibility} of systems of discrete relations and
for constructing {canonical decompositions} of discrete
relations.
\par
Applying the technique described above to some
cellular automata --- a particular case of discrete
relations --- we obtained a number of 
results. The most interesting among them,
in our opinion,  is the demonstration of how the presence
of non-trivial {proper consequences} may determine
the global behavior of an automaton.
\par
We suggest an algorithmic approach ---  based on discrete symmetry
  analysis and implemented in C --- for construction and investigation of
 discrete dynamical models --- {deterministic}, {mesoscopic} 
 and {quantum}.
We hope that our approach can be used in various practical applications,
such as, for example,  simulation of nanostructures with 
nontrivial symmetry properties.  
\par
We demonstrate that soliton-like moving structures --- 
like ``{spaceships}'' in cellular automata, ``traveling waves'' 
in mathematical physics and ``generalized coherent states'' in quantum physics 
--- arise inevitably in  {deterministic} dynamical 
systems whose symmetry group splits the set of states
into finite number of group orbits. 
\par 
We formulate the gauge principle in the form most suitable for  
discrete and finite systems.
We also propose a method --- based on introduction of unitary gauge connection 
  of a special kind --- for quantizing discrete systems and construct 
  simple models for studying properties of suggested quantization.
\par
We show that if unitary operators describing dynamics of finite quantum system 
form finite group, then the system can be embedded into a classical system with a simple
behavior.
We hope that discrete and finite background allowing comprehensive study 
may lead 
to deeper understanding of the quantum behavior and its connection 
with symmetries of systems.
\par
To study more complicated models we are developing C programs based on 
computer algebra and  computational group theory methods.
\section*{Acknowledgments}
This work was supported by the grant 10-01-00200 from the Russian
Foundation for Basic Research and by the grant 3810.2010.2 from the
Ministry of Education and Science of the Russian Federation.

\label{lastpage-01}


\begin{thebibliography}{99}
\bibitem{Verlinde}
Verlinde, E.P. (2010). \textit{On the Origin of Gravity and the Laws of Newton,}\\ arXiv:1001.0785
\bibitem{tHooft}
't Hooft, G. (1993). \textit{Dimensional reduction in quantum gravity,}\\ 
Utrecht preprint THU-93/26;
gr-qc/9310006.
\bibitem{KornyakOC}
Kornyak, V.V. (2005).
On Compatibility of Discrete Relations,
\emph{Lect. Notes Comp. Sci.} \textbf{3718},  Springer-Verlag Berlin
Heidelberg,  272--284.
\bibitem{KornyakDR}
Kornyak, V.V. (2006).
Discrete Relations On Abstract Simplicial Complexes,
\emph{Programming and Computer Software}. \textbf{32}, No 2, 84--89.
\bibitem{KornyakCASC06}
Kornyak, V.V. (2006).
Cellular Automata with Symmetric Local Rules,
\emph{Lect. Notes Comp. Sci.} \textbf{4194},  Springer-Verlag Berlin
Heidelberg, 240--250.
\bibitem{Kornyak08}
Kornyak, V.V. (2008). Discrete Dynamical Systems with Symmetries:
Computer Analysis,
\emph{Programming and Computer Software}. \textbf{34}, No 2, 84--94.
\bibitem{KornyakCASC09}
Kornyak, V.V. (2009).
Discrete Dynamics: Gauge Invariance and Quantization,
\emph{Lect. Notes Comp. Sci.} \textbf{5743},  Springer-Verlag Berlin
Heidelberg, 180--194.
\bibitem{Holt}
Holt, D.F.;~Eick, B.;~O'Brien,~E.~A. (2005). \emph{Handbook of Computational Group Theory}.
Chapman \& Hall/CRC Press
\bibitem{Seiberg}
Seiberg, N. (2006). Emergent Spacetime. Rapporteur talk at the 23rd Solvay Conference in
Physics, December, 2005. hep-th/0601234.
\bibitem{McKay}
McKay,~B.D. (1981). Practical Graph Isomporphism. \emph{Congressus Numerantium} \textbf{30}, 45--87, 
http://cs.anu.edu.au/~bdm/nauty/PGI
\bibitem{Klein}
Klein, F. (1884). \emph{Vorlesungen \"{u}ber das Ikosaeder}. Leipzig:
Teubner. Translated to Russian under the title \emph{Lektsii ob ikosaedre
i reshenii uravnenii pyatoi stepeni}, Moscow:
Nauka, 1989.
\bibitem{Kirillov}
Kirillov, A.A. (1976). \textit{Elements of the Theory of Representations.}
 Springer-Verlag, Berlin-New York.
\bibitem{Lidl}
Lidl,~R.; Niederreiter, H. (1983).
\emph{Finite Fields;} Reading, Mass.: Addison-Wesley.
\bibitem{Hilton}
Hilton, P.;  Wayle,~S. (1960). \emph{Homology Theory: An Introduction to Algebraic Topology,} 
Cambrige Univ. Press.  
\bibitem{lives-site}
http://psoup.math.wisc.edu/mcell/rullex\_life.html 
\bibitem{Wolfram}
Wolfram, S. (2002).  \emph{A New Kind of Science;} Wolfram Media, Inc
\bibitem{site}
http://atlas.wolfram.com/01/01/
\bibitem{tHooft99}
't Hooft, G. (1999).
Quantum Gravity as a Dissipative Deterministic System. 
SPIN-1999/07, gr-qc/9903084; 
\emph{Class. Quant. Grav. }\textbf{16}, 3263 (1999);
 also published in: \textit{Fundamental Interactions: 
from symmetries to black holes} (Conference held on 
the occasion of the ``Em\'{e}ritat'' of Fran\c{c}ois Englert, 
24-27 March 1999, ed. by J.-M. Fr\`{e}re et al, 
Univ. Libre de Bruxelles, Belgium, 221--240. 
\bibitem{tHooft06}
't Hooft, G. (2006).
The mathematical basis for deterministic quantum mechanics. 
ITP-UU-06/14, SPIN-06/12, quant-ph/0604008, 1--17. 
\bibitem{Imry02}
Imry, Y. (2002). \emph{Introduction to Mesoscopic Physics 
(Mesoscopic Physics and Nanotechnology, 2).} Oxford University Press, USA,  256 p.
\bibitem{Gross01}
Gross, D.H.E. (2001). \emph{Microcanonical thermodynamics: Phase transitions in ``Small'' Systems.} 
World Scientific, Singapore,  269 p.
\bibitem{Gross04}
Gross, D.H.E. (2004).  A New Thermodynamics from Nuclei to Stars. \emph{Entropy}, \textbf{6}, 
158–-179 
\bibitem{GrossVotyakov}
Gross, D.H.E.; Votyakov, E.V. (2000). Phase Transitions in ``Small'' Systems. \emph{Eur. Phys. J. B}, 
\textbf{15}, 115–-126.  
\bibitem{IspolatovCohen}
Ispolatov, I.; Cohen, E. G. D. (2001). On First-order Phase Transitions in Microcanonical 
and Canonical Non-extensive Systems, \emph{Physica A} \textbf{295}, 475--487.
\bibitem{Oeckl}
Oeckl, R. (2005).
\textit{Discrete Gauge Theory (From Lattices to TQPT)}. 
Imperial College Press, London.
\bibitem{Feynman}
Feynman R.P.; Hibbs A.R. (1965). \textit{Quantum Mechanics and Path Integrals.} McGraw-Hill.
\bibitem{Hall}
Marshall Hall, Jr. (1959). \textit{The Theory of Groups.} Macmillan Co., New York. 
\bibitem{Serre}
Serre, J.-P. (1977). \textit{Linear Representations of Finite Groups.} Springer-Verlag.
\bibitem{KornyakFQM10}
Kornyak, V.V. (2010).~
\textit{Finite Quantum Models: Constructive Approach to Description 
of Quantum  Behavior}, 
http://arxiv.org/abs/1010.3370
\end{thebibliography}
\end{document}